\documentclass[aps,pra,singlecollumn,10pt,nofootinbib,floatfix] {revtex4-2}
 
\usepackage{xcolor}

\usepackage[utf8]{inputenc} 
\usepackage[T1]{fontenc}
\usepackage{csquotes}

\usepackage{sqrcaps}

\usepackage{ulem}

\usepackage{graphicx}
\usepackage{dcolumn}
\usepackage{bm}
\usepackage{epstopdf}

 \usepackage{lineno}
 \usepackage{amssymb}
\usepackage{hyperref}

\usepackage{yfonts}

\usepackage{calligra}
\usepackage{mathrsfs}

\newcommand{\comment}[1]{}

\newcommand{\rmi}{{\rm i}}

\newcommand{\rmA}{{\rm A\,}}
\newcommand{\rmB}{{\rm B\,}}
\newcommand{\rmC}{{\rm C}}
\newcommand{\rmF}{{\rm F}}
\newcommand{\rmM}{{\rm M\,}}
\newcommand{\rmS}{{\rm S\,}}

\newcommand{\rmMB}{{\rm MB\,}}
\newcommand{\rmSA}{{\rm SA\,}}
\newcommand{\rmSB}{{\rm SB\,}}
\newcommand{\rmSM}{{\rm SM}}

\newcommand{\hatri}{\hat r_i}
\newcommand{\hatRih}{\hat R_i^h}

\newcommand{\up}{\uparrow}
\newcommand{\down}{\downarrow}

\newcommand{\myskip}[1]{}   
\newcommand{\mijnskip}[1]{}   

\newcommand{\scriptD}{\hat{\cal D}}

\newcommand{\BEQ}{\begin{eqnarray}}      
\newcommand{\EEQ}{\end{eqnarray}}      
\newcommand{\BEA}{\begin{eqnarray}}      
\newcommand{\EEA}{\end{eqnarray}}      
      
\renewcommand{\d}{{\rm d}}      
\newcommand{\half}{\frac{1}{2}}

\newcommand{\cD}{{\cal D}}
\newcommand{\cE}{{\cal E}}

\newcommand{\cH}{{\cal H}}

\newcommand{\cN}{{\cal N}}

 \newcommand{\cP}{\mathbb{P}}

\newcommand{\cR}{{\cal R}}

\newcommand{\ti}{t_{\rm i}}
\newcommand{\tf}{t_{\rm f}}
\newcommand{\tr}{{\rm tr}\,}
\newcommand{\Tr}{{\rm Tr}\,}

\newcommand{\eq}{{\rm eq}}

\newcommand{\vv}{{\bf v}}

\begin{document}

\nolinenumbers

 \title{Teaching ideal quantum measurement, from dynamics to interpretation \\  \hspace{3mm}
Enseigner les mesures quantiques id\'eales, de la dynamique \`a l’interpr\'etation}
 
 
 \author{Armen E. Allahverdyan$^1$, Roger Balian$^2$ and Theo M. Nieuwenhuizen$^ {3}$}

\address{$^1$ Yerevan Physics Institute, Alikhanian Brothers Street 2, Yerevan 375036, Armenia \\
              $^2$ Institut de Physique Th\'eorique, CEA Saclay, 91191 Gif-sur-Yvette  cedex, France \\
              $^3$ Institute for Theoretical Physics,  Science Park 904, 1098 XH Amsterdam, The Netherlands}

 \vspace{1cm}

\begin{abstract}
\noindent\noindent \\ {\bf Abstract} \\ 
We present a graduate course on ideal measurements, analyzed as dynamical processes of interaction between the tested system S and an apparatus A, described by quantum statistical mechanics. The apparatus A=M+B involves a macroscopic measuring device M and a bath B. The requirements for ideality of the measurement allow us to specify the Hamiltonian of the isolated compound system S+M+B. The resulting dynamical equations may be solved for simple models. Conservation laws are shown to entail two independent relaxation mechanisms: truncation and registration. Approximations, justified by the large size of M and of B, are needed. The final density matrix $\hat\cD(\tf)$ of S+A has an equilibrium form. It describes globally the outcome of a large set of runs of the measurement. The measurement problem, i.e., extracting physical properties of individual runs from $\hat\cD(\tf)$, then arises due to the ambiguity of its splitting into parts associated with subsets of runs. To deal with this ambiguity, we postulate that each run ends up with a distinct pointer value $A_i$ of the macroscopic M. This is compatible with the principles of quantum mechanics. Born’s rule then arises from the conservation law for the tested observable; it expresses the frequency of occurrence of the final indications $A_i$ of M in terms of the initial state of S.  Von Neumann’s reduction amounts to updating of information due to selection of $A_i$. We advocate the terms $q$-probabilities and $q$-correlations when analyzing measurements of non-commuting observables.
 These ideas may be adapted to different types of courses.
 \\   \\  \\
 {\bf R{\'e}sum\'e}      \\
 On pr\'esente un cours doctoral sur les mesures id\'eales, processus dynamiques couplant le syst\`eme test\'e S et un appareil A analys\'es en m\'ecanique statistique quantique. 
 Cet appareil A=M+B comprend un dispositif de mesure macroscopique M et un bain B. 
 Les conditions requises pour l’id\'ealit\'e de la mesure impliquent une forme sp\'ecifique du Hamiltonien du syst\`eme composite isol\'e S+M+B. 
Les equations dynamiques r\'esultantes sont solubles pour des mod\`eles simples. Les lois de conservation engendrent deux m\'ecanismes de relaxation ind\'ependants, la troncature et l’enregistrement. Des approximations, justifi\'ees par la grande taille de M et de B, sont n\'ecessaires. La matrice densit\'e finale $\hat\cD(\tf)$ de S+A a une forme d’\'equilibre. Elle d\'ecrit globalement l’issue d’un large ensemble de processus similaires. Le probl\`eme de la mesure, extraire de $\hat\cD(\tf)$ des propri\'et\'es physiques de processus individuels, provient ici de l’impossibilit\'e de le scinder sans ambigu\"it\'e en parties d\'ecrivant des sous-ensembles de processus. On l\`eve cette ambigu\"it\'e en postulant que chaque mesure individuelle aboutit \`a une valeur distincte $A_i$ du pointeur macroscopique. Ceci est compatible avec les principes de la m\'ecanique quantique. La r\`egle de Born r\'esulte alors de la loi de conservation de l’observable mesur\'ee; elle exprime la fr\'equence de chaque indication finale $A_i$  de M en termes de l’\'etat initial de S. La r\'eduction de von Neumann appara\^it comme une mise \`a jour de l’information r\'esultant de la s\'election d’un r\'esultat $A_i$. On pr\'econise l’emploi des termes $q$-probabilit\'es ou $q$-corr\'elations lors de l’analyse de mesures d’observables non-commutatives. Ces id\'ees peuvent être adapt\'ees \`a divers types de cours.\footnote{Comptes Rendus. Physique,  Volume 25 (2024), pp. 251-287. DOI: 10.5802/crphys.180}
 \end{abstract}

 \maketitle

\vspace{5mm}
 
 \vspace{5mm}

Mots clefs: mesures quantiques id\'eales, $q$-probabilit\'es, dynamique syst\`eme-appareil, probl\`eme de la mesure, r\`egle de Born, r\'eduction de von Neumann, 
interpr\'etation minimaliste, contextualit\'e
 

\vspace{2cm}

 \tableofcontents

\renewcommand{\thesection}{\arabic{section}}
\renewcommand{\thesubsection}{\arabic{subsection}}
\renewcommand{\thesubsubsection}{\arabic{subsubsection}}
 \section{ Introduction}
\renewcommand{\thesection}{\arabic{section}.}
\renewcommand{\thesubsection}{\thesection.\arabic{subsection}}
\renewcommand{\thesubsubsection}{\thesubsection.\arabic{subsubsection}}

\makeatletter
\renewcommand{\p@subsection}{}
\renewcommand{\p@subsubsection}{}
\makeatother


Most courses of quantum mechanics include lectures on ideal measurements, defined as thought processes characterized by Born's rule 
and von Neumann's reduction of the state. Students may be puzzled by this topic, as these two properties are often presented as separate basic principles, while reduction may {seem to} contradict
 Schr\"odinger's equation.  \cite{carroll2022addressing,mermin2022there}. 
Clarifying the status of such properties is pedagogically important, since they greatly contribute to the understanding of quantum mechanics. 
This issue is a current concern of teachers, who tackle it from many different viewpoints. 
We propose here a consistent approach that has been tested on graduate students and that might also inspire courses at less advanced levels. 
It relies on the standard quantum dynamics for the macroscopic apparatus coupled to the tested system, complemented with a postulate about the apparatus which affords a solution of the measurement problem.

\renewcommand{\thesection}{\arabic{section}.}
\renewcommand{\thesubsection}{\thesection\arabic{subsection}}
\renewcommand{\thesubsubsection}{\thesubsection.\arabic{subsubsection}}

\subsection{Interpretation, ideal measurement, apparatus and dynamics}
\renewcommand{\thesubsection}{\arabic{subsection}.}

\label{sec1.1}

The quantum formalism encodes in a mathematical language our knowledge about the systems under study, and affords efficient predictions. However, this description is abstract, and relating it to physical reality requires an interpretation. The literature devoted to this matter is immense, often subtle, and encompasses many different viewpoints (for reviews, see 
 \cite{wheeler2014quantum,busch1996quantum,peres1997quantum,van1988ten,styer2002nine,de2006foundations,laloe2019we,wiseman2009quantum}). 
 Students should get acquainted with the subject, which includes understanding of Born's rule and von Neumann's reduction, but some specific approach 
 suitable for teaching must be selected.

By ``interpretation'', we will mean a complete theoretical analysis of 
thought processes that relate the quantum formalism to experimental facts (§ \ref{sec1.2} (c)). 
 The study of measurements thus provides insight on quantum theory. It helps to grasp how physical knowledge 
 about a tested object can be derived from its mathematical representation.

As we wish to fully work out the theory of a measurement process, we should consider a sufficiently simple situation. A current approach, initiated by 
Jordan and von Neumann  \cite{von2018mathematical}, consists in focusing on ``ideal quantum measurements''. They are dynamical processes through which 
a single observable of the system S is tested by an apparatus A, and such that S is not perturbed by its interaction with A if it initially lies in an eigenstate of this observable. 
 They involve many runs, starting alike but possibly yielding one or another among several different outcomes\footnote{Measurement theory then provides the relative frequency 
 of occurrence of each possible outcome. If one considers a single process, this frequency is identified with the probability of occurrence of the considered result.\label{newfna}}.
They will be defined more precisely in § \ref{sec3.1} and § \ref{sec5.1}. 

As an illustrative example, we  consider the so-called Curie-Weiss model \cite {allahverdyan2003curie,allahverdyan2013understanding}.
There, we imagine measuring the $z$-component $s_z$ of S, a spin $\half$, by letting S interact with an Ising ferromagnet A initially in a metastable paramagnetic state. If $s_z=+1$ (or $-1$), the process should keep $s_z$ unchanged, while triggering the pointer, 
which is here the $z$-component $M_z$ of the macroscopic total magnetization of A, so that it finally takes  the value $M_z=+M_\rmF$ (or  $M_z=-M_\rmF$).
 We wish to explore the situation in which the initial state of S is arbitrary.
 A detailed introduction to this magnetic model and its complete treatment are presented in ref.     \cite{allahverdyan2013understanding}, 
 which also includes a wide review of other measurement models.
 
A general model describing arbitrary ideal measurements is specified in § \ref{sec3}. Laboratory measurements rarely satisfy these features 
 \cite{clerk2010introduction,braginsky1996quantum,svensson2013pedagogical}, and we should regard ideal measurements,
also called quantum non-demolishing measurements,  as thought experiments. 
However, their analysis provides ideas relevant to real measurements.

We designate as ``apparatus'' A the {\it complete} experimental setup coupled to S.
The idealized system A includes a measuring device M which models the whole experimental machinery, including detectors that afford 
registration and reading,  and a thermal bath B,  often called ``environment''' or ``context'', which will ensure equilibrium at the issue of the process. 
The compound system S+A, which contains the bath B, is isolated, which will allow to describe its evolution through the Liouville--von Neumann equation of motion.

Consistency requires that the joint evolution of S+A obeys standard quantum mechanics, both S and A being treated as quantum objects. 
However, A is {\it macroscopic}, involving a huge number of degrees of freedom. Moreover, the dynamical process is {\it irreversible} on the timescales of physical interest. 
These two reasons exclude a description by means of pure states.
Since the macroscopic apparatus A cannot be described in detail, we must rely on quantum {\it statistical}  mechanics and use density operators. 
This also allows to treat a mixed state for S,  the most general case.
The final density operator\footnote{Quantum mechanical operators are indicated by a hat.} $\hat\cD(\tf)$  of S+A
 thus determined is expected to express how the indication of M reflects a property of S. 
 However, it is still a formal object, and it will remain to face interpretation issues (§ \ref{sec1.3} and § \ref{sec6}).

\renewcommand{\thesubsection}{\arabic{subsection}}
\renewcommand{\thesubsection}{\thesection\arabic{subsection}}
\renewcommand{\thesubsubsection}{\the-subsection.\arabic{subsubsection}}
\subsection{The measurement problem}

\label{sec1.3}

Consider a given observable $\hat s$ of the system S with eigenvalues $s_i$ and suppose that S has been prepared in such a way that $\hat s$ takes the 
(non-degenerate) value $s_1$ among its set of eigenvalues $s_i$. This property characterizes the fact that S is in a pure state represented by the eigenstate 
$|\psi_1\rangle$ of $\hat s$ (or by the density operator $\hat r_1=|\psi_1\rangle\langle\psi_1|$). Then, in an ideal measurement of $\hat s$, the joint dynamics of S+A, after interaction of S with A, leaves S unchanged while M is modified. Its pointer variable reaches, among the possible macroscopic indications $A_i$, 
the value $A_1$ associated with $s_1$, revealing the initial property of S (§ \ref{sec3.2.3}).
Thus, the quantum system M should be chosen so as to involve a {\it measurement basis} (§ \ref{sec3.2.1}), 
the elements of which are in one-to-one correspondence with the eigenvalues $s_i$,  this correspondence being ensured by the form 
of the coupling between S and M (§ \ref{sec3.2.2}).

Non-classical features appear if the state $\hat r$ of S, whether pure or not, is not an eigenstate of $\hat s$. Then, if a measurement of $\hat s$ is performed on different samples S, all prepared in the same initial state $\hat r(0)$ at the initial time $\ti=0$, a subsequent ideal measurement of $\hat s$ is expected to provide, at the final time $t=t_{\rm f}$ of each individual run, different 
possible outcomes $i$. The set $\cE$ of runs can thus be separated into subsets $\cE_i$. Both S and  A, in particular the pointer of M, are modified. The subset $\cE_i$  gathers the runs with outcome $A_i$ of the pointer of M, which occur with some probability $p_i$,  and for which  S ends up in a final state $\hat r_i$ (§ \ref{sec3.1} and \ref{sec5.1}). 
The celebrated {\it measurement problem}  \cite{laloe2019we} arises from the apparent contradiction between the uniqueness of the initial state and the non-uniqueness of the final state. 

In fact, what we call ``quantum state'', whether pure or not, is in general an abstract mathematical object which allows us to make probabilistic predictions about the outcomes of any 
experiment that will test the object under study, for arbitrary observables. It encodes everything that we can say about this object, 
and is related only indirectly to its physical properties (§ \ref{sec2}). It is thus irreducibly probabilistic: even the above pure state $|\psi_1\rangle$ 
is a probabilistic mathematical object with regard to  observables other than $\hat s$, as their measurement on different samples all described by $|\psi_1\rangle$ 
will yield different answers, the probabilities of which are determined by $|\psi_1\rangle$. 

Thus, within the ``frequentist'' interpretation of probabilities, a quantum state refers to a large set of similar systems $^{\ref{newfna}}$. 
In particular, the density operator $\cD(t)$ of S+A associated with an ideal quantum measurement process does not describe only a single operation, 
but a large set $\cE$ of similarly prepared individual runs\footnote{Note that Bell 
 \cite{bell1988physics,bell1989against}   advocated to replace the word ``measurement'' by ``experiment''; he argued that speaking of ``measurement'' incites to 
refer to some pre-existing property of the sole object, irrespective of the apparatus.}.

The density operator $\hat\cD(\tf)$ describing S+A at the issue of the interaction process, for the full set of runs, can be determined in the framework 
of quantum statistical mechanics. One has to solve the equation of motion for $\hat\cD(t)$, with a known initial condition on $\hat\cD(0)$ (§ \ref{sec4}).

We wish to derive from $\hat\cD(\tf)$ the features of {\it individual runs of a measurement}, in particular to understand von Neumann's reduction, 
which characterizes the different final states $\hat r_i$ of S that may occur in the various runs, and Born's rule, which expresses their relative number $p_i$. 
However, as will be discussed in § \ref{sec5.2}, the object $\hat\cD(\tf)$ cannot yet be interpreted physically, although it contains the $\hat r_i$'s and $p_i$'s 
as ingredients. It is a {\it mathematical}  tool describing, {\it only globally}, the outcome of the {\it large set of runs} of the ideal measurement. 
Its sole knowledge is not sufficient to ensure the very existence of subsets for which the pointer of the apparatus takes some macroscopic value $A_i$ 
and to  extract the required information about {\it individual runs}. This peculiarity of density operators in the standard quantum formalism 
is the mathematical form taken here of the measurement problem. It is stated more precisely in § \ref{sec5} below.

To face this mismatch, we will supplement the standard formal quantum rules with an idea issued from macroscopic experience. Namely, we acknowledge 
that {\it each individual run} of the measurement produces a {\it single macroscopic indication of the pointer of the apparatus} which may be read 
or registered (§ \ref{sec5.3}). This statement, although intuitive, is not a consequence of the mere quantum principles and has a macroscopic nature. 
We will accept its inclusion in quantum theory  as a {\it postulate} (motivated by the macroscopic size of M),
and also show its {\it full consistency} with the quantum formalism (§ \ref{sec6.1}). We shall then be in position to interpret $\hat\cD(\tf)$,
splitting it into meaningful terms (§ \ref{sec5.4}) and to derive therefrom the expected properties of ideal measurements (§ \ref{sec6}).

\renewcommand{\thesubsection}{\thesection\arabic{subsection}}
\renewcommand{\thesubsubsection}{\thesubsection.\arabic{subsubsection}}
\subsection{Overview}
\renewcommand{\thesubsection}{\arabic{subsection}.}

\label{sec1.4}

We present below the scheme of various courses about the dynamics and the interpretation of ideal measurements which have been delivered by one or another of us at Universities or Summer Schools. The full approach is suited to advanced students, having some knowledge of quantum statistical mechanics, but basic ideas deserve to be retained in more elementary courses. The length of the text is partly due to inclusion of details and remarks issued from questions of students. 
Complementary exercises are proposed; some may be inspired by numerous models which describe the joint evolution of S+A.
The present text, somewhat sketchy and abstract, is intended for teachers. They may construct a course suited to their audience by finding inspiration in refs. 
 \cite{paris2012modern,allahverdyan2013understanding},
 which contain more explanations, a review of solvable models and a detailed treatment of the Curie--Weiss model, with illustrative figures that should help the understanding.

We first resume the {\it principles} of quantum statistical mechanics in a form suited to this problem (§ 2). 
In particular, we regard a ``state'' as an abstract tool gathering our whole knowledge (Bohr's ``catalog of knowledge'')
of a large set of systems, in view of making probabilistic predictions. 
It has some features of ordinary probability distributions but should not be confused with them. We also recall its dynamical equation and the definition of von Neumann’s entropy. 

We then recall the expected properties of ideal measurements and define a {\it general model} for them (§ \ref{sec3.1}). The compound object under study involves the tested system S and the apparatus A, in which we distinguish the measuring device M and the thermal bath B. Relying on the conditions needed to ensure that the interaction process between S and A behaves as an ideal measurement, we specify in § \ref{sec3.2} the required form of the Hamiltonian of S+M+B.
(This tripartite model, suited for our purpose, is not the only one that can be imagined.)

In § \ref{sec4.1} we write the expression of the initial state $\hat\cD(0)$ of S+M+B and the form taken by the equations of motion, taking into account the conservation laws. The final state $\hat\cD(\tf)$ should be determined dynamically from these data.  Students acquainted with statistical mechanics may 
study, for various solvable models,  the relaxation of $\hat\cD(t)$ to find the explicit form of $\hat\cD(\tf)$ (§ \ref{sec4.2} and Appendices \ref{secA}, 
\ref{secB}, \ref{secC})  \cite{allahverdyan2013understanding}. Suited exercises are proposed. For less advanced courses, we propose a more expedient 
and general procedure, relying on a {\it thermodynamic approach} (§ 4.3).\footnote{This ``thermodynamic approach'' differs from Neumaier's ``thermal 
interpretation'' of quantum mechanics  \cite{neumaier2019coherent}, although both analyses rely on the consideration of macroscopic variables and 
the use of quantum statistical mechanics.\label{footn4}}
Namely, keeping aside the bath and describing by the marginal state $\hat D={\rm tr}_\rmB \hat\cD$ the system S+M weakly coupled to B, we first derive the general form of the {\it thermodynamic equilibrium} states $\hat D_\eq$ of S+M, obtained by maximizing von Neumann’s entropy. 
We then identify the dynamics of the measurement process with a relaxation towards thermodynamic equilibrium. Thus, the final state
$\hat D(\tf)$ belongs to this equilibrium class, and we determine it by using conservation laws.

As indicated above (§ \ref{sec1.3}) and explained in § \ref{sec5}, we then have to face the measurement problem: the final density operator $\hat D(\tf)$ found for S+M for the {\it full set} $\cE$ of runs cannot readily be separated into parts that would describe the result of {\it individual runs} of $\cE$ (§ \ref{sec5.2}). But by relying on experimental evidence, we postulate that the value of the pointer is macroscopically well-defined for each run (§ \ref{sec5.3}). This allows us to split $\hat D(\tf)$ into physically meaningful parts (§$\,\,$\ref{sec5.4}). 

We can thus derive the properties of the outcomes of ideal measurements (§ \ref{sec6}), stressing that their physical interpretation depends crucially on the apparatus (§ \ref{sec6.4}). 
Although $\hat\cD(\tf)$ does not have the same properties as a standard probability distribution (§ \ref{sec2.1.3}), the coefficients $p_i$ entering $\hat\cD(\tf)$ will come out as 
ordinary probabilities $p_i$, understood as relative frequencies of the possible outcomes $A_i$ for the pointer at the final time $\tf$. Born’s rule, which expresses these $p_i$’s 
in terms of the initial state of S, will appear as a consequence of the joint dynamics and of the conservation laws (§ \ref{sec6.3}). Von Neumann’s reduction will be interpreted 
as an updating of the state of S based on the knowledge of  $A_i$ (§ \ref{sec6.4}).

We conclude (§ \ref{sec7}) by suggesting which parts of this text seem adapted to such or such level of teaching, from introductory courses of quantum mechanics to doctoral studies.

\renewcommand{\thesection}{\arabic{section}}
\section{Quantum principles}
\renewcommand{\thesection}{\arabic{section}.}

\label{sec2} 

We noted in § \ref{sec1.1} that the analysis of measurement processes requires quantum statistical mechanics. Among the many formulations of quantum theory, we recall its principles in a form suited to that purpose  \cite{david2015formalisms}. When being taught, the rules resumed below should be illustrated by examples.

\renewcommand{\thesubsection}{\thesection\arabic{subsection}}
\renewcommand{\thesubsubsection}{\thesubsection.\arabic{subsubsection}}
\subsection{Mathematical structures}
\renewcommand{\thesubsection}{\arabic{subsection}.}

\label{sec2.1}

The mathematical structure of the quantum formalism involves two dual spaces, that of ``observables'' and that of ``states'', abstract objects related only indirectly to the corresponding physical entities (§ \ref{sec1.2} , point (a))  \cite{darrigol2015some}. 

\renewcommand{\thesubsection}{\thesection\arabic{subsection}}
\renewcommand{\thesubsubsection}{\thesubsection.\arabic{subsubsection}}
\subsubsection{Observables}
 \renewcommand{\thesubsubsection}{\arabic{subsubsection}}

\label{sec2.1.1}

 The physical quantities attached to a given type of system are called ``{\it observables}'', and mathematically represented by operators $\hat O$ belonging to a non-commutative algebra.\footnote{Mathematically, observables are self-adjoint elements of a C$^\ast$-algebra, a structure which involves an addition $\hat A_1+\hat A_2$, a product $\lambda\hat A$ by complex numbers $\lambda$, a non-commutative multiplication $\hat A_1 \hat A_2$, a unit element $\hat I$, an involution that defines conjugate pairs $\hat A \leftrightarrow\hat A^\dagger$, a norm $| | \hat A| | $, together with consistency rules such as associativity $\hat A_1(\hat A_2 \hat A_3)=(\hat A_1 \hat A_2)\hat A_3$ and distributivity $\hat A_1(\hat A_2+\hat A_3)= \hat A_1\hat A_2+ \hat A_1\hat A_3$. \label{fn2.1}}
For the systems with a finite number of degrees of freedom that we are considering, the observables can be represented by Hermitean matrices ($\hat O =\hat O^\dagger$) in a Hilbert space $\cH$. Their eigenvalues are expected to be tested by inference from the macroscopic outcomes of measurements.

\renewcommand{\thesubsection}{\thesection\arabic{subsection}}
\renewcommand{\thesubsubsection}{\thesubsection.\arabic{subsubsection}}
\subsubsection{States and sets of systems} 

\label{sec2.1.2}

As indicated in § \ref{sec1.2}, quantum theory does not describe in general individual systems, but an {\it infinitely large collection}
$\cE$ of similarly prepared systems, real or possibly virtual. For such a collection, the quantum formalism associates, at each time, 
a numerical value $\langle \hat O\rangle$  with each physical quantity represented by an observable $\hat O$. 
This correspondence, which is a mathematical representation of a ``{\it quantum state}'', is formally defined as a linear mapping $\hat O \mapsto  \langle \hat O\rangle$  (for $\hat O =\hat O^\dagger$), such that $\langle \hat I\rangle  = 1$ for the unit observable $\hat I$, that $\langle \hat O\rangle$  is real and that $\langle \hat O^2\rangle  - \langle \hat O\rangle ^2 \ge 0$. For systems with a finite number of degrees of freedom, it can be expressed in the form  
\BEQ \label{eq2.1}
\langle \hat O\rangle  =\Tr \hat \cD \hat O,			
\EEQ
by a {\it density operator} $\hat \cD$ in the (self-dual) Hilbert space $\cH$. In a given basis, $\hat \cD$ is represented by means of a {\it density matrix} which is Hermitean, non-negative and normalized.\footnote{We shall indifferently use the terms ``state'', ``density operator'' and ``density matrix''.} Pure states $|\psi\rangle$,  whose density operators are projectors $\hat \cD =|\psi \rangle \langle \psi| $
on a one-dimensional subspace of $\cH$, appear as extreme points of the convex set of states $\hat \cD$ (see end of Appendix \ref{secD}).
 Contrary to what it often done in elementary courses, we do not assign them a more fundamental status nor a special interpretation.

The above formalism presents some analogy with the standard probability theory. Observables $\hat O$ look like random variables, states $\hat\cD$ 
like probability distributions, and (\ref{eq2.1}) like expectation values. However, this analogy is purely formal (§ \ref{sec1.3} and § \ref{sec2.1.4}). 
Not only mathematical properties but also basic concepts are different for the two theories. 
We shall rely here on the interpretation of ordinary probabilities  \cite {mises1957probability} as  {\it frequencies of occurrence of exclusive events}
(such as heads or tails), which take place among an infinitely large set of individual draws, whereas quantum theory involves {\it non-exclusive} properties, 
such as the polarization of a spin $\half$ along $+z$ or along $+x$.  Hence, the overall properties of a quantum state do not comply with ordinary probability 
theory. Accordingly, it will be essential below (§ \ref{sec5}) to carefully distinguish the {\it infinitely large collection} $\cE$ of individual systems, 
or of individual runs of an experiment, from a ``statistical ensemble'', governed by the ordinary probability theory.
 Such a set $\cE$, described by a quantum state $\hat\cD$, is just a collection of similar elements,  while a {\it statistical ensemble} is an infinitely large 
 set of systems which display exclusive events that {\it may be counted} to define probabilities.
``Observables'' and ``states'' are mathematical symbols without yet a physical interpretation. They will give rise to genuine probabilities 
(in the sense of relative frequencies) at the issue of measurement theory (§ \ref{sec6.1}), but only for the single tested observable.

\renewcommand{\thesubsection}{\thesection\arabic{subsection}}
\renewcommand{\thesubsubsection}{\thesubsection.\arabic{subsubsection}}
\subsubsection{Physical meaning of a quantum state}
 \renewcommand{\thesubsubsection}{\arabic{subsubsection}}

\label{sec2.1.3}

Students should not be misled by the term, ``state'', currently used to designate the mathematical object $\hat\cD$, and difficult to avoid when teaching. 
Its meaning should be made clear. Whether ``pure'' or not, the so-called ``quantum state of a system'' is not an intrinsic physical property of an individual system. 
We advise to avoid the expression ``the wave function of the particle’’.
As stressed by Bohr  \cite{pais1991niels,plotnitsky2006reading}, a ``state'' $\hat\cD$ is just a {\it mathematical tool} which encompasses 
``{\it what we can say'' about a large set}  of systems, allowing us to make {\it probabilistic predictions} about future experiments.
As indicated above, we adhere here to the interpretation of a probability as the relative {\it frequency} of occurrence of each type of event, in a large set of events  \cite{plotnitsky2006reading}.
However, the results of our approach will eventually apply to  {\it individual} processes, if one relies on the Laplace – Bayes interpretation where a probability appears as a {\it likelihood} 
(see  \cite{allahverdyan2013understanding},  § 10.1.3).  

One should never forget that a quantum state, represented by a density operator, a density matrix, a pure state or a wave function, 
is an {\it abstract object} encoding our {\it best knowledge about the set of systems} under consideration.\footnote{\label{fnbestknowledge} The {\it best knowledge}
may even differ for different observers, who rely on their available data to assign different states to a set of systems (§ \ref{sec2.3}). 
More complete knowledge leads to more optimal predictions for future experiments, as will be exemplified in  §  \ref{sec6.3}.\label{fn6}}

\renewcommand{\thesubsection}{\thesection\arabic{subsection}}
\renewcommand{\thesubsubsection}{\thesubsection.\arabic{subsubsection}}
\subsubsection{Probabilities versus $q$-probabilities}
 \renewcommand{\thesubsubsection}{\arabic{subsubsection}}

\label{sec2.1.4}

Consider the special case of a single observable $\hat O \equiv \sum_\alpha  \omega_\alpha \hat\pi_\alpha$ of a system S (with eigenvalues $\omega_\alpha$ and eigen-projectors $\hat\pi_\alpha$) and of a density operator $\hat\cD_\alpha$ belonging to the eigenspace $\alpha$ of $\hat O$  (i. e., constrained to satisfy
 $\hat\cD_\alpha\equiv \hat\pi_\alpha\hat\cD_\alpha\hat\pi_\alpha$ or $\Tr \hat\cD_\alpha \hat\pi_\alpha=1$). Then, the number $\langle\hat O\rangle=\Tr \hat\cD_\alpha \hat O= \omega_\alpha$ (with $\langle\hat O^2\rangle -\langle\hat O\rangle^2=0$) is identified with the {\it physical value} $\omega_\alpha$ taken by the quantity $\hat O$  in the considered state. As will be proven in § \ref{sec3.2.3}, subsequent measurements of $\hat O$  keep $\hat\cD_\alpha$ 
 unchanged and will again yield $\omega_\alpha$.

Still considering a single observable $\hat O$  but an arbitrary density operator $\hat\cD$, we will analyze below an ideal measurement process testing $\hat O$, where S interacts with an apparatus A, both S and A being treated as quantum objects. This will lead us (§ \ref{sec6.2}) to interpret, but {\it only for this single observable}, the quantities $p_\alpha$ formally defined by $p_\alpha = \Tr \hat\cD \hat\pi_\alpha$ as relative frequencies of occurrence of the values $\omega_\alpha$ and hence $\langle\hat O\rangle = \Tr\hat\cD \hat O  = \sum_\alpha  p_\alpha \omega_\alpha$ as an ordinary expectation value.

However, such an interpretation fails as soon as we measure  {\it several non-commuting} observables. For instance, for a $q$-bit (whose observables are the Pauli matrices $\hat \sigma_{1,2,3}$ of a spin $\half$), the occurrences of the four possible values $s_z = \pm 1$, $s_x = \pm 1$ do not constitute a sample space of four exclusive events to which probabilities might be assigned, in contrast to the two alternative values of an ordinary bit. If we consider the 4 projectors $\hat\pi _\lambda$  associated with the events $s_z = \pm 1$, $s_x = \pm 1$, we cannot regard the 4 quantities $\Tr\hat\cD \hat\pi _\lambda$  as probabilities of exclusive events.

To discuss such questions, the traditional vocabulary is a source of confusion. Whenever we deal with non-commuting observables, and already in introductory 
courses, we advocate to name ``$q$-{\it probabilities}'' rather than ``probabilities'' the formal quantities $\langle\hat\pi \rangle=\Tr\hat\cD \hat\pi$  (on the model of $q$-bit 
versus bit).\footnote{The mathematical structure of these so-called ``quantum probabilities'', which differs from that of ordinary probabilities, will not be presented here. 
The Boolean algebra describing elementary events in ordinary probability theory is replaced by the ``lattice structure” of the set of non-commutative projectors, 
with which the above formal quantities are associated. The standard interpretations of probabilities in terms of elementary physical events are then lost.}
Likewise, as observables are directly expressed in terms of projectors,
the quantities $\langle\hat O\rangle = \Tr\hat\cD\hat O$  should be called ``$q$-{\it expectation values}''. For a {\it single product} $\hat O= \hat O_1 \hat O_2$ of commuting observables,  
$\langle\hat O_1 \hat O_2 \rangle$ may be regarded as a correlation, but for several products 
$\hat O= \hat O_1 \hat O_2$, $\hat O'= \hat O_1' \hat O_2'$, $\hat O''= \hat O_1'' \hat O_2''$, $\cdots$, such that $\hat O$, $\hat O'$, $\hat O''$, $\cdots$, 
do not commute, the set $\langle\hat O_1\hat O_2\rangle$, $\langle\hat O_1'\hat O_2'\rangle$, $\langle\hat O_1''\hat O_2''\rangle$, $\cdots$,
should systematically be called ``$q$-{\it correlations}'', as they do not have {\it together} the properties of correlations. For instance they violate Bell's inequalities  \cite{bell1966problem,bell1988physics,aspect1982experimental,aspect1982experimentaltest} that constrain ordinary correlations. 
This would help the students to grasp quantum properties which {\it violate rules of ordinary probabilities} (Bell's inequalities  \cite{bell1966problem,bell1988physics,aspect1982experimental,aspect1982experimentaltest}) or even seem to violate logic (GHZ paradox  \cite{greenberger1989going}). 
In such cases, apparently paradoxical quantum properties arise from combining several {\it formal} $q$-correlations 
that involve non-commuting observables. In fact, the determination of such different $q$-correlations requires measurements performed on 
{\it different samples} by {\it different apparatuses}, so that we cannot interpret them {\it together} as physical correlations (§ \ref{sec6.4}). 
The misleading term ``correlations'' suggests fake paradoxes. Likewise, the two ``$q$-{\it variances}'' entering Heisenberg's inequality 
$\langle{\hat O}_1^2\rangle \langle{\hat O}_2^2\rangle \ge \frac{1}{4}|\langle[{\hat O}_1 , {\hat O}_2]\rangle|^2 $, which has a quantum nature, 
are neither ``uncertainties'' nor ordinary ``variances'', as  testing them refers to incompatible measurements involving different apparatuses.

In fact, even for a single observable, it is of pedagogical interest to denote formal quantities computed from a quantum state as $q$-probabilities, 
since they reach a physical meaning only after recording through measurement. Indeed, 
we will show in § \ref{sec6} how the ideal measurement of a {\it single observable} of S allows us to interpret the $q$-probabilities associated with its eigen-projectors as ordinary probabilities (in the sense of relative frequencies). This property relies on the interaction of S with a macroscopic {\it dedicated apparatus}. Measuring two non-commuting observables requires different apparatuses, the outcomes of which cannot be put together to produce ordinary probabilities.

 \renewcommand{\thesubsection}{\thesection\arabic{subsection}}
\renewcommand{\thesubsubsection}{\thesubsection.\arabic{subsubsection}}
\subsection{Basic laws}

\label{sec2.2}

The laws of quantum theory (§ \ref{sec1.2}, point (b)) are expressed by mathematical operations on density operators.

\renewcommand{\thesubsection}{\thesection\arabic{subsection}}
\renewcommand{\thesubsubsection}{\thesubsection.\arabic{subsubsection}}
\subsubsection{Evolution}

\label{sec2.2.1}

 The time dependence of the density operator $\hat \cD(t)$ of an {\it isolated} system (such as S+A) is generated by a Hamiltonian $\hat H  $ which depends on the nature of the system. It is governed by the Liouville--von Neumann dynamical equation
 \BEQ \label{eq2.2}
    \rmi\hbar  \frac{\d\hat \cD (t)}{\d t} = [\hat H  , \hat \cD (t)], 		
\EEQ
reducing to the Schr\"odinger equation $\rmi\hbar\,  \d|\psi \rangle /\d t = \hat H |\psi \rangle$  for pure states 
$|\psi\rangle$ represented by a projector $\hat \cD =|\psi \rangle \langle \psi| $.

\renewcommand{\thesubsection}{\thesection\arabic{subsection}}
\renewcommand{\thesubsubsection}{\thesubsection.\arabic{subsubsection}}
\subsubsection{Subsystems} 

\label{sec2.2.2}

For a compound system described by the state $\hat \cD$, here the union of the tested system S and the apparatus A, the marginal density matrices $\hat r$ and $\hat \cR$ which describe the subsystems S and A are obtained from $\hat \cD$ by taking the {\it partial traces} $\hat r= {\rm tr}_A \hat \cD$ and $\hat \cR= {\rm tr}_\rmS  \hat \cD$ over A and S, respectively.\footnote{
We denote by $\Tr$ a scalar obtained by taking the trace of an operator over the full associated Hilbert space, and by $\tr$ an operator obtained by taking a partial trace over a compound Hilbert space. For instance, if $\hat X$ acts in the compound Hilbert space $\cH_\rmSA=\cH_\rmS   \otimes \cH_\rmA  $ , ${\rm tr}_\rmA  \hat X$ is an operator acting in the Hilbert space $\cH_\rmS  $ and we have ${\rm Tr}\, \hat X \equiv {\rm Tr}_\rmSA \hat X = {\rm Tr}_\rmS   ({\rm tr}_\rmA  \hat X) = {\rm Tr}_\rmA  ({\rm tr}_\rmS   \hat X)$.}

\renewcommand{\thesubsection}{\thesection\arabic{subsection}}
\renewcommand{\thesubsubsection}{\thesubsection.\arabic{subsubsection}}
\subsubsection{ Merging of sets}

\label{sec2.2.3}

Consider two large sets $\cE_1$ and $\cE_2$ of similar systems,  described in the Hilbert space $\cH$ by given states $\hat\cD_1$ and $\hat\cD_2$, respectively. 
Merging $\cN_1$ elements extracted from $\cE_1$ and $\cN_2$ elements extracted from $\cE_2$ produces a new set $\cE$ gathering the $\cN=\cN_1+\cN_2$ elements. 
The  compound set $\cE$ is described (as in ordinary probability theory) by the weighted sum
\BEQ \label{eq2.3}
\cN\hat \cD = \cN_1 \hat \cD_1 + \cN_2 \hat \cD_2 .             
\EEQ

However, the {\it converse is wrong}, as will be demonstrated in § \ref{sec5.2} and Appendix \ref{secD}. Given a density operator $\hat \cD$ that represents some set $\cE$ of systems, we cannot infer from a {\it mathematical} decomposition $\hat \cD = \lambda_1\hat \cD_1 + \lambda_2\hat \cD_2$ as a weighted sum that $\hat \cD_1$ and $\hat \cD_2$ represent {\it physical} subsets of $\cE$, although $\hat \cD_1$ and $\hat \cD_2$ have the properties of density operators (§ \ref{sec2.1.2}). This specifically quantum property must be faced to solve the measurement problem (§ \ref{sec5}).

\renewcommand{\thesubsection}{\thesection\arabic{subsection}}
\renewcommand{\thesubsubsection}{\thesubsection.\arabic{subsubsection}}
\subsubsection{Von Neumann entropy}

\label{sec2.2.4}

In agreement with the definition (\ref{eq2.1}) of a state $\hat \cD$ as a catalogue of probabilistic knowledge about a large collection of systems, 
one associates with it the von Neumann entropy\footnote{We take units in which the Boltzmann constant equals  $k=1$.},
 \BEQ \label{eq2.4}
S(\hat \cD) \equiv  -\Tr \hat \cD\ln \hat \cD, 		
\EEQ
a number which measures (in dimensionless units) the {\it amount of information} which is missing when only $\hat \cD$ is specified  \cite{balian2005information}.
Roughly speaking, it can also be viewed as a measure of disorder.
It will be used to assign a state to a partially known set of systems (§ \ref{sec2.3} and § \ref{sec4.3.1}) and to estimate 
the loss of information carried by the density operator of S+A, due to the irreversibility of the measurement process (Appendix \ref{secC}) 
as well as the gain of information about S (§ \ref{sec6.3}).

\renewcommand{\thesubsection}{\thesection\arabic{subsection}}
\renewcommand{\thesubsubsection}{\thesubsection.\arabic{subsubsection}}
\subsection{Preparation}
\label{sec2.3}

The preparation of a large collection $\cE$ of systems at a given time, in view of performing future experiments, is physically ensured by an apparatus 
which imposes some data to $\cE$, and theoretically accounted for by assigning to $\cE$ a density operator $\hat \cD$. 
We shall encounter this issue in several circumstances. 

A preparation is characterized by some {\it given data}, imposed to all the systems of the set  $\cE$. Each one is implemented as a 
{\it constraint} on the state $\hat \cD$ that we wish to specify. For instance, if a preparation apparatus which controls the physical quantity 
$\hat O \equiv \sum_ \alpha  \omega_\alpha  \hat \pi_\alpha$ imposes that it always takes the value $\omega_\alpha$,  the state $\hat \cD$ 
must satisfy the constraint $\Tr  \hat \cD\,\hat \pi_\alpha  =1$ (§ \ref{sec2.1.4}), meaning that the value $\omega_\alpha$  is determined by 
$\scriptD$ with certainty. 
(We expect in this case that the outcome of a measurement of  $\hat O$ will be $\omega_\alpha$, see § \ref{sec3.2.3}.)
 In the limit of very large systems, we may also deal with {\it macroscopic data}, defined not exactly but within a negligible uncertainty. 
 For instance, the macroscopic energy $E$ of a thermal bath (with Hamiltonian $\hat H$) is given by the $q$-expectation $\Tr \hat \cD\hat H = E$,  
 where $E$ lies within the dense spectrum of $\hat H$, the large size of the bath allowing the $q$-fluctuation of energy around $E$ to be negligible in 
 relative value at the macroscopic scale. The macroscopic indication of the pointer (§ \ref{sec3.1}) will have the same features: dense spectrum, negligible $q$-fluctuation.

When such constraints are imposed through the control of a complete set of commuting observables, they fully determine the state $\hat \cD$ (which is then pure). Otherwise, and especially for macroscopic systems, various density operators $\hat \cD$ are compatible with the known data, and one should select among them the {\it least biased} one by means of a probabilistic argument. This choice relies on the {\it maximum entropy criterion} 
 \cite{elsasser1937quantum,jaynes1957information12,balian2007microphysics12}
based on the identification of the von Neumann entropy (\ref{eq2.4}) as a measure of randomness: The sought density operator $\hat \cD$ is the one that maximizes $-\Tr \hat \cD\ln  \hat \cD$ under the given constraints\label{fn2.2}.\footnote{This use of the von Neumann entropy in the determination of the least biased state $\hat \cD$ can be justified by taking as a prior distribution 
in the space of states the one that is invariant under unitary transformations, then by relying on the indifference 
 (or equiprobability) principle of Laplace. This procedure produces the same result $\hat \cD$ as the maximum entropy criterion  \cite{balian1987equiprobability}.}
 In case the data are expectation values that do not change over time, this procedure yields a {\it thermodynamic equilibrium state} (§ \ref{sec4.3}); if the sole constraint is $\Tr \hat \cD\hat H  =E$, it yields a canonical equilibrium state $\hat \cD \sim \exp (- \beta  \hat H  )$ (Eqs. (\ref{eq3.2}), (\ref{eq3.3})). 

\renewcommand{\thesection}{\arabic{section}}
\section{A model for ideal measurements and its formal implementation}
\renewcommand{\thesection}{\arabic{section}.}

 \label{sec3}
  
  The above principles cannot be given any direct  and complete interpretation, although some have already be given an interpretation 
  (beginning of §§ \ref{sec2.1.4},   \ref{sec2.2.4} and  \ref{sec2.3}).  What we will interpret below, in agreement with contextuality, is only the result of measurements.
  
 In the continuation, we will first analyse in the above quantum formalism the dynamics of idealized experiments that may be regarded as ideal measurements (§ \ref{sec4}). 
 We will then search for a physical interpretation of the outcome of the process, in order to finally clarify the physical meaning of Born's rule and of von 
 Neumann’s reduction which characterize ideal measurements (§ \ref{sec6}). 
 We gather here beforehand their specific physical features, sometimes subtle but needed, and build a general model on which our subsequent study of the measurement process will rely. These various features are exemplified by the Curie-Weiss model 
  \cite{allahverdyan2013understanding}, which can be used throughout to illustrate them.

\renewcommand{\thesubsection}{\thesection\arabic{subsection}}
\renewcommand{\thesubsubsection}{\thesubsection.\arabic{subsubsection}}
\subsection{Physical requirements}

\label{sec3.1}

The purpose is to  {\it explore the properties of an observable}\label{fn3.0}\footnote{For instance, for a spin $\half$, 
$\hat s_z= \hat \pi _\uparrow -\hat \pi_\downarrow$  with $\hat\pi_\uparrow=\left(\!\!\begin{array}{ll} 1 \,\,0 \\ 0 \,\, 0 \end{array}\!\!\right)$
and $\hat\pi_\downarrow=\left(\!\!\begin{array}{ll} 0 \,\,0 \\ 0 \,\, 1 \end{array}\!\!\right)$ has the eigenvalues $\pm 1$. } 
 $\hat s \equiv  \sum_i s_i \hat \pi_i$ (with eigenvalues $s_i$ and eigen-projectors $\hat \pi_i$) of a system S described by a given density operator  $\hat r (0)$ at the initial time $\ti=0$, by letting S interact with a dedicated apparatus. Moreover, the ideal measurement process as defined in  § \ref{sec1.1}  should {\it perturb} S {\it as little as possible} (§ \ref{sec3.2.2}). To this aim, S is dynamically coupled to an idealized apparatus A dedicated to this task. We distinguish two parts A=M+B in this apparatus. The part M is a {\it measuring mechanism} including a {\it macroscopic pointer} that should eventually display outcomes $A_i$ in one-to-one correspondence with $s_i$ at the final time $\tf$ of the process. A {\it thermal bath} B weakly coupled to M will contribute to the equilibration of S+M.

This  idealized experiment consists of {\it many repeated runs}.\footnote{\label{fn11} This large number of runs is not only a standard experimental 
feature, but is also theoretically needed because the quantum formalism cannot describe an individual measurement (§ \ref{sec2.1.2}). 
In fact, Born's probabilities refer to a large set of runs; they represent the relative number of runs that have provided each possible outcome. }
The large set $\cE$ of compound quantum systems S+A = S+M+B involved in these runs is described at the time $t$ by a state $\hat \cD(t)$ in the Hilbert space $\cH=\cH_\rmS  \otimes \cH_\rmM \otimes\cH_\rmB $. We shall denote by $\hat r$, $\hat R_\rmM $, $\hat R_\rmB $, $\hat \cR$ and $\hat D$ the marginal density operators (§ \ref{sec2.2.2}) of the subsystems S, M, B, A=M+B and S+M in their respective Hilbert spaces $\cH_\rmS $, $\cH_\rmM $, $\cH_\rmB $, $\cH_A=\cH_\rmMB$ and $\cH_\rmSM$, e. g., $\hat D={\rm tr}_\rmB \hat \cD$, $\hat r={\rm tr}_\rmM \hat D$. As M and B must be {\it macroscopic}, 
states such as $\hat \cD(t)$, $\hat R_\rmM (t)$ or $\hat D(t)$ {\it cannot be pure}\footnote{A pure state, or a wave function, cannot be assigned 
to a macroscopic system like an apparatus.  Characterizing a piece of material by a pure state would require control of an unrealistic number 
of variables}, a fact already stressed in §   \ref{sec1.1}.
The {\it Hamiltonian} $\hat H$ governing the dynamics of the isolated system S+M+B can be decomposed as\footnote{More precisely, as $\hat H$ 
acts in the full Hilbert space $\cH=\cH_\rmS\otimes\cH_\rmA$ and $\hat H_\rmS$ in the Hilbert space $\cH_\rmS$ of S, 
 $\hat H_\rmS$ in (\ref{eq3.1}) is a short hand for $\hat H_\rmS\otimes \hat I_\rmA$, where $\hat I_\rmA$ is the unit operator in the Hilbert space 
 $\cH_\rmA$. And likewise for all other terms in (\ref{eq3.1}). We shall use the same simplified notation in the continuation.}
\BEQ
\label{eq3.1}
	\hat H   = \hat H  _\rmS  + \hat H _\rmA + \hat H_\rmSM, \hspace{1mm}   \hat H _\rmA = \hat H  _\rmM  + \hat H  _\rmB  + \hat H  _\rmMB,  
\EEQ
into parts associated with the three subsystems and with their interactions. 

The measuring device M should have the following physical property. For each run, it should end up at the time $\tf$ in one or another of some {\it stable states} $\hat R_i$, 
which are often called ``{\it pointer states}'',  characterized by the {\it macroscopic indication} $A_i$ of the pointer, which can be read or registered within a proper time window. 
The pointer variable is represented quantum mechanically by an observable $\hat A$ of M, and each outcome by a narrow region of the spectrum of $\hat A$ centred 
around the value $A_i$. Each state $\hat R_i$ should thus exhibit $A_i$ as the $q$-expectation value $A_i = {\rm tr}_\rmM  \hat R_i \hat A$ of $\hat A$, while the 
relative fluctuation of $\hat A$ in $\hat R_i$ should be negligible for large M.


One wishes the interaction process between M and S to produce at the final time a {\it one-to-one correspondence} between the outcome $A_i$ of $\hat A$ and the 
measured eigenvalue $s_i$ of $\hat s$. In particular, if the tested system S is initially prepared in a state $\hat r(0)$ such that $\hat s$ takes the value $s_i$ with certainty, 
the subsequent interaction $\hat H  _\rmSM$ between M and S should {\it trigger the evolution of}  M towards the pointer state $\hat R_i$, despite the macroscopic size of M and 
the smallness of S, which should remain unaffected. This is possible in case M had beforehand been prepared in a {\it metastable initial state} $\hat R_\rmM (0)$ which 
may decay towards one or another stable state $\hat R_i$, depending on a small perturbation. 
We will implement this property in the general model of ideal measurement considered below.

The {\it bath} B will contribute to the relaxation of M through the weak coupling $\hat H  _\rmMB$, which allows dumping of free energy from M to B during the process. 
This bath, at temperature $T$, described by the thermal equilibrium state (end of § \ref{sec2.3}),
\BEQ \label{eq3.2}
\hat R_\rmB  \sim  \exp\left(-\frac{H _\rmB }{T}\right), 		
\EEQ
is sufficiently large so that its marginal density operator ${\rm tr}_\rmSM\hat D(t) =\hat R_\rmB (t)\simeq  \hat R_\rmB $ remains 
nearly unchanged throughout the process. 
We recall that the initial state of the apparatus A,  $\hat R_\rmM \otimes\hat R_\rmB$, is a {\it mixed} state.

\renewcommand{\thesubsection}{\thesection\arabic{subsection}}
\renewcommand{\thesubsubsection}{\thesubsection.\arabic{subsubsection}}
\subsection{Features of the Hamiltonian} 

\label{sec3.2}

Some specific conditions must be satisfied by M so that it can be used as a measuring device, and by the interactions between S, M and B 
so that their coupled dynamics can be regarded as an ideal measurement process. We construct here the general form (\ref{eq3.1}) of the 
Hamiltonian of the compound system S+M+B needed to ensure the required properties of ideal measurements.

Let us first specify the properties of the Hamiltonian of the apparatus A=M+B alone required so that M acts as an ideal measuring device. 

.

\renewcommand{\thesubsection}{\thesection\arabic{subsection}}
\renewcommand{\thesubsubsection}{\thesubsection.\arabic{subsubsection}}
\subsubsection{Apparatus alone}

\label{sec3.2.1}

 The existence for M of {\it several} possible final pointer states $\hat R_i$ as well as the existence of an initial {\it metastable} state $\hat R_\rmM  (0)$ 
 imply that M is macroscopic and should be ensured by some specific form of its own Hamiltonian $\hat H_\rmM$. These features also require that the system M is macroscopic. 
 The different states $\hat R_i$ must stand on the same footing to avoid bias in the measurement, so that the macroscopic energies $U_i = {\rm tr}_\rmM  \hat R_i \hat H  _\rmM $ 
 should be the same as well as the entropies $S (\hat R_i)$. A simple means of ensuring such properties consists in assuming that the Hamiltonian $\hat H_\rmM $ possesses 
 some {\it symmetry which is broken at the temperature} $T$ of the bath.\footnote{The group of transformations  that keeps $\hat H_\rmM$ invariant under permutation of the indices $i$, 
 which is the parity $+z\leftrightarrow -z$ for the Curie--Weiss model with spin $\half$, has been non-trivially generalised to higher spins  \cite{nieuwenhuizen2022models}.}

The implementation of this property may be illustrated for students by the Curie--Weiss model, studied in detail in  \cite{allahverdyan2013understanding}. In that thought experiment, 
the apparatus A alone is simulated by a piece of ferromagnetic material, modelled as a system M of $N$ 
spins $\half$ with Ising interaction, weakly coupled to a phonon bath B at temperature $T$. 
The Hamiltonian $\hat H_\rmM $ is invariant under the exchange  $+z\leftrightarrow -z$, and the pointer observable $\hat A$ of M is the total magnetization $\hat M_z$ along 
the $z$-direction, proportional to $N$. If $T$ lies below the Curie temperature $T_\rmC$, the paramagnetic initial state $\hat R_\rmM (0)$ of M (with $\langle \hat M_z\rangle =0$) 
is metastable, and two ferromagnetic equilibrium states $\hat R_i$ exist, namely $\hat R_\Uparrow$  and $\hat R_\Downarrow$  which are characterized by the values 
$\langle\hat M_z\rangle=+ M_\rmF$ and $-M_\rmF$, respectively.\footnote{We refer to the directions $+z$ and $-z$ by using double arrows $\Uparrow$ and $\Downarrow$ for M 
(e.g. $\hat R_\Uparrow$ or  $\hat R_\Uparrow$)  and single arrows $\uparrow$ and $\downarrow$ for S (e.g.  $\hat\pi_\uparrow$).}
Beyond the application to measurements, the selection of one state $\hat R_i$ of M or the other is usually 
achieved by considering a situation in which a weak magnetic field directed towards either $+z$ or $-z$ is applied to M.
A source term $\hat h_\Uparrow=-g\hat M_z$ (or $\hat h_\Downarrow=g\hat M_z$)  which breaks the symmetry $+z\leftrightarrow -z$ is thus added to $\hat H_\rmM $. 
The state $\hat R_\Uparrow$ (or $\hat R_\Downarrow$) is obtained by letting $g$ become small {\it after} the
 number $N$ of magnetic 
spins is made large\label{fn3.1}.\footnote{Strictly  speaking,  maximization of the von Neumann entropy $S(\hat R_\rmM )$ for a given energy (§ 2.3) yields the single thermodynamic equilibrium state 
$\hat R_{\eq} \propto \exp (-\hat H  _\rmM /T)$, with entropy $S_\eq$. The states $\hat R_\Uparrow$ and $\hat R_\Downarrow$ have an entropy 
$S(\hat R_\Uparrow) = S(\hat R_\Downarrow)$ nearly equal to $S_\eq$ for large $N$, as $\hat R_\eq$ is very close to $\half  (\hat R_\Uparrow  + \hat R_\Downarrow )$. 
The state $\hat R_\Uparrow$ ($\hat R_\Downarrow$) is located in a subspace $\cH_\Uparrow$ ($\cH_\Downarrow$) of $\cH_\rmM$ spanned by a set of eigenvectors 
of $\hat M_z$ associated with eigenvalues that lie around $+M_\rmF$ ($-M_\rmF$), within a small interval of relative size $1/\sqrt{N}$. The dimension $\cN\sim\exp S_\eq$ 
of $\cH_\Uparrow$ ($\cH_\Downarrow$)  is exponentially large in $N$. Then, although the function $S(\hat R_\rmM)$ is concave, its maximum $S_\eq$ is nearly flat. 
The entropy $S_\eq$ is practically reached for the huge set of density matrices $\hat R_{\rm maxent}$ located in the space  $\cH_\Uparrow\oplus\cH_\Downarrow$ and 
having a dimension of order $\cN$. However, if the Hamiltonian $\hat H_\rmM$ contains weak interactions which allow transitions between neighboring eigenstates of 
$\hat M_z$, the states $\hat R_{\rm maxent}$ relax towards one or another of the dynamically stable states $\hat R_{\lambda} = \lambda \hat R_\Uparrow +(1- \lambda )\hat R_\Downarrow$
 ($0\le\lambda\le1$), the value of  $\langle \hat M_z\rangle $ remaining nearly fixed. (This may be checked as an exercise by working out the ``poly-microcanonical'' relaxation mechanism
 \cite{allahverdyan2017sub}.) Mathematically, the states $\hat R_{\lambda }$, including the pointer states $\hat R_\Uparrow$ and $\hat R_\Downarrow$, are not rigorously stable and relax 
 for finite $N$ towards $\hat R_\eq$ (with $\lambda=1/2$), but extremely slowly. In fact, the relaxation time is exponentially large in $N$ and all states $\hat R_{\lambda}$ can physically 
 be regarded as fully stable. More generally, for an arbitrary ideal measurement, the very long lifetime of the states $\hat R_i$ allows the outcome $A_i$ to be read not much later 
 than the end $\tf$ of the process.\label{footn3.1}}


Likewise, in the general case, we wish to formally define each expected possible final equilibrium state $\hat R_i$ of M.
As explained in footnote \ref{footn3.1}, we cannot do that by considering the sole Hamiltonian $\hat H_\rmM$. 
To this aim, we first add as above a small {\it source term}  $\hat h_i$ to its Hamiltonian $\hat H_\rmM$. This produces the single equilibrium state\footnote{The quantities $U_i$, 
$S_i$ and  $Z_i= Z=\Tr \exp [- (\hat H_\rmM +\hat h_i) /T] = \exp (S_i-U_i/T)$ are the same for all $i$.}
\BEQ \label{eq3.3}
 \hatRih =\frac{1}{Z_i} \exp \left(- \frac{\hat H_\rmM  +\hat h_i}{T}\right).             
\EEQ
(This introduction of operators $\hat h_i$ may look artificial, but its relevance to measurement will be explained in § \ref{sec3.2.2}.) 
Denoting by $A^h _i$ the $q$-expectation value of the pointer operator $\hat A$ in the state $\hat R^h_i$, we select the operators $\hat h_i$ in such a way that the corresponding  $q$-fluctuations
 are small compared to the differences between the values $A^h _i$. We then let each small source $\hat h_i$ decrease, remembering that M is very large.
For a suitable choice of $\hat H  _\rmM $ and sufficiently low $T$, the states $\hatRih$ shift only slightly, so as to reach {\it different limits} $\hat R_i$, 
as in the case of materials with broken invariance. The macroscopic pointer variable $A_i$ appears as the limit of the $q$-expectation value 
\BEQ \label{eq3.4}
		{\rm Tr}_\rmM  \hatRih \hat A  \to  {\rm Tr}_\rmM  \hat R_i \hat A = A_i .     	
\EEQ
(For the Curie-Weiss model, $A_i=M_{zi}=\pm M_\rmF$.) 

Concerning the full apparatus A = M + B, we do not need here to specify the Hamiltonian $\hat H  _\rmB $ of the bath B nor its weak interaction $\hat H_\rmMB$ with M, as we admit below (§ \ref{sec4.3}) that this coupling ensures the relaxation of S+M towards thermodynamic equilibrium. 
The operators $\hat H_\rmB$ and $\hat H_\rmMB$ are defined for models  \cite{allahverdyan2013understanding} 
in which the dynamics of this relaxation can be treated by statistical 
mechanics (§ \ref{sec4.2}). As the pointer variable must change during this process, $\hat H_\rmMB$ should not commute with $\hat A$.
(In the Curie-Weiss model, $\hat H_\rmMB$ induces transitions between eigenstates of $\hat H_\rmM$ which involve neighboring eigenvalues of 
the magnetization per spin  $\hat m=\hat M/N$.

\renewcommand{\thesubsection}{\thesection\arabic{subsection}}
\renewcommand{\thesubsubsection}{\thesubsection.\arabic{subsubsection}}
\subsubsection{Coupling between the system and the apparatus}

\label{sec3.2.2}

We noted (§ \ref{sec3.1}) that the evolution of the macroscopic apparatus M towards one $\hat R_i$ or another is driven by its coupling $\hat H  _\rmSM$ with the tested system S. 
In return, S is also generally perturbed by its coupling with M. We have defined (§ \ref{sec1.1}) an ideal measurement as a process for which this perturbation is minimal 
(i. e.,  it is absent if $\hat r=\hat\pi_i\hat r\hat\pi_i$). Let us determine the general form of $\hat H_\rmSM$ that fulfils this requirement. 

First, in an ideal experiment the dynamics in the Heisenberg picture
should not affect the tested observable $\hat s$. This is expressed by the commutation $[\hat H , \hat \pi_i] = 0$ of the full Hamiltonian with the eigen-projectors $\hat \pi_i$ of $\hat s$, i.e., $\hat \pi_i\hat H  \hat \pi_j=0$ for $ i\neq  j$\label{fn3.3}.\footnote{\label{fn14}Strictly speaking, $\hat \pi_i$ acts in the Hilbert space $\cH_\rmS $. In such equations pertaining to the space $\cH$, $\hat \pi_i$ is meant as $\hat \pi_i \otimes\hat I_\rmA$, where $\hat I_\rmA$ is the unit operator in the space $\cH_\rmA$. Likewise, in § 4.3.1, $\hat x_\alpha $ stands for $\hat x_\alpha \otimes\hat I_\rmM $.}
This conservation law is an essential feature of ideal measurements, already stressed by Wigner and Yanase  \cite{wigner1973analysis}.
 Moreover, in case eigenvalues of $\hat s$ are degenerate, dynamics should {\it not affect} the observables $\hat x$ of S {\it physically compatible with} $\hat s$.  This compatibility is expressed by the commutation of $\hat x$ with all eigen-projectors $\hat \pi_i$ of $\hat s$. 
 Hence, besides the $\hat \pi_i$'s, the conserved quantities are the set
\BEQ \label{eq3.5}
\hat x = \sum_i \hat \pi_i \hat y \hat \pi_i ,                            
\EEQ
where $\hat y$ is any arbitrary Hermitean operator in $\cH_\rmS $. (For example, if S is a pair of spins or $q$-bits S$'$, S$''$ and if the measured quantity
 is $\hat s'_z$, the observables $\hat x$ are all linear combinations of $\hat s'_z$, $\hat s''_x$, $\hat s''_y$, $\hat s''_z$, $\hat s'_z\hat s''_x$, 
 $\hat s'_z\hat s''_y$ and $\hat s'_z\hat s''_z$.) The required {\it conservation laws} $[\hat H  , \hat x] = 0$ in the space $\cH$ then entail that 
 $[\hat \pi_i\hat H  \hat \pi_i, \hat y]=0$,   for each $i$ and for any $y$ in the space $\cH_\rmS$. 
 This implies that the operator $\hat\pi_i \hat H_\rmSM \hat\pi_i$ should have the form $\hat\pi_i \otimes \hat K_i$, where $\hat K_i$ acts only in the Hilbert 
 space 
 $\cH_\rmM$\footnote{Likewise,  $\hat H_\rmS $  must have the form $\hat H_\rmS  = \sum_i \varepsilon_i \hat \pi_i$. Such a contribution $\hat H_\rmS  $ to $\hat H$  
 has the sole effect of multiplying the off-diagonal blocks $\hat \pi_i \hat \cD(t)\hat \pi_j$ by the phase factor $\exp i(\varepsilon_j - \varepsilon_i)t/\hbar$. 
 As these blocks eventually disappear at $t = \tf$, we set $\hat H_\rmS  = 0$, in which case S evolves only through its coupling to A.}.
 In the special case when $\hat  r(0) = \hat\pi_i \hat r(0) \hat\pi_i $, S is unaffected  by the dynamics in the Schr\"odinger picture
 while M is governed by the effective Hamiltonian $\hat H_\rmM+\hat K_i$. 
As we wish then the pointer to take the macroscopic value $A_i$, the operator $\hat K_i$ should have the form of a source $\hat h_i$ introduced in § \ref{sec3.2.1}  to define the state (\ref{eq3.3}).
 Altogether, since $\hat\pi_i \hat H_\rmSM \hat\pi_j=0$ for $i \neq j$, the interaction between S and M should have the form
\BEQ \label{eq3.6}
\hat H_\rmSM = \sum_i \hat \pi_i \otimes \hat h_ i  .   
\EEQ
With such an interaction, we expect that, if $\hat r(0) =\hat\pi_i\hat r(0)\hat\pi_i$, M reaches at the end of the joint relaxation process of S+M the marginal state given by (\ref{eq3.3}). 
If afterwards the coupling between S and M is switched off, and if $\hat h_i$ is sufficiently small, we expect M to reach $\hat R_i$ as in (\ref{eq3.4}).

For instance, in the Curie-Weiss model of measurement, we saw in § \ref{sec3.2.1} that the magnet M can be driven towards 
 states close to the ferromagnetic states  $\hat R_\Uparrow$ or $\hat R_\Downarrow$ by sources of the form $\hat h_\Uparrow = -g\hat M_z$ 
 or $\hat h_\Downarrow = +g\hat M_z$.   Thus, if the tested spin S is coupled to M through 
$\hat H_\rmSM =\hat\pi_\uparrow\otimes \hat h_\Uparrow+\hat\pi_\downarrow\otimes \hat h_\Downarrow=-g\hat s_z\otimes\hat M_z$, 
with a small but finite $g$,   M is triggered by $\hat h_\Uparrow$ if S is oriented in the $+z$-direction, or by $\hat h_\Downarrow$ if S lies in the $-z$-direction.   
In case the initial paramagnetic state $\hat R(0)$  is partly stabilized by a free energy barrier, $g$ must be {\it large enough} to overcome this barrier 
and allow the relaxation towards one of the stable ferromagnetic states.

\renewcommand{\thesubsection}{\thesection\arabic{subsection}}
\renewcommand{\thesubsubsection}{\thesubsection.\arabic{subsubsection}}
\subsubsection{A non-perturbing measurement}

\label{sec3.2.3}

As defined in § \ref{sec1.1} and recalled in § \ref{sec3.1}, an ideal measurement requires M to {\it end up in the state} $\hat R_i$ {\it if} $\hat s$ {\it takes the value} 
$s_i$ in the initial state $\hat r(0)$ of S.  We  focus here on the dynamics of S + A for an initial state $\hat r(0)$ of S satisfying $\hat r(0) \equiv  \hat \pi_i \hat r(0)\hat \pi_i$, 
a case already considered in § \ref{sec3.2.2} to construct  $\hat H_\rmSM$. (If the eigenvalue $s_i$ of $\hat s$ is non-degenerate, $\hat r(0)=\hat\pi_i$.) 
The solution of the Liouville--von Neumann equation (\ref{eq2.2}) then behaves as $\cD(t)=\hat r(0) \otimes \cR(t)$, where 
\BEQ \label{eq3.7}
\rmi\hbar  \frac{\d\hat \cR(t)}{\d t} = [\hat H_\rmM  +\hat h_i+\hat H_\rmB +\hat H_\rmMB, \hat \cR(t)] .
\EEQ
In this special case, the system S remains unaffected and the coupling (\ref{eq3.6}) has the sole effect of yielding the dynamical equation (\ref{eq3.7}) for the apparatus. 
This equation is the same as if M+B were alone (§ \ref{sec3.2.1}), but with $\hat H_\rmM  +\hat h_i$ as proper Hamiltonian of M,
the operator $\hat h_i$ issued from $\hat H_\rmSM$ behaving as a source that favours the outcome $A_i$ as in § \ref{sec3.2.1}. 
The contributions $\hat H_\rmB +\hat H_\rmMB$ of the bath are expected to drive the marginal state $\hat R(t)={\rm tr}_\rmB \hat \cR(t)$ of M towards the equilibrium state (\ref{eq3.3})
 $\hat R_i^h$ expressed by (\ref{eq3.3}). Then, at the very end of the process, S and A can be separated,  $\hat H_\rmSM$ is switched off. 
 The measuring device M then evolves from $\hat R_i^h$ to the neighbouring final state $\hat R_i$  as in  (\ref{eq3.4}),  leaving practically unchanged the indication $A_i$ of the pointer.

 \renewcommand{\thesection}{\arabic{section}}
 \renewcommand{\thesubsection}{\arabic{subsection}}
 \renewcommand{\thesubsubsection}{\arabic{subsubsection}}
\section{Relaxation for the full set of runs}
 \renewcommand{\thesection}{\arabic{section}.}
 \renewcommand{\thesubsection}{\arabic{subsection}.}
 \renewcommand{\thesubsubsection}{\arabic{subsubsection}.}

\label{sec4}

Any large set of isolated quantum systems is described by a well-defined density operator that evolves deterministically, according to (\ref{eq2.2}). 
For ideal measurements, this idea applies to the large set $\cE$ of compound systems S+A. We now determine the change of their density operator 
$\hat\cD(t)$ between the initial and final times $\ti=0$ and $\tf$ of the process.

\renewcommand{\thesubsection}{\thesection\arabic{subsection}}
\renewcommand{\thesubsubsection}{\thesubsection.\arabic{subsubsection}}
\subsection{Initial state and equations of motion}

\label{sec4.1}

We have beforehand to assign to S+A, along the lines of § \ref{sec2.3}, a density operator $\hat \cD(0)$ which accounts for our knowledge of S+A at the initial time $t = 0$ of the process. The tested system S is supposed to have been prepared earlier in a given state $\hat r(0)$, and the measuring device M in a ``ready'' {\it metastable} state $\hat R_\rmM (0)$ characterized by some macroscopic data. The large bath B lies in the equilibrium state (\ref{eq3.2}) at temperature $T$. (For instance, in the Curie--Weiss model, M is initially set at $T>T_\rmC$ in a paramagnetic state with $\langle \hat M_z\rangle =0$.) The {\it initial} state of the compound system S+M+B is therefore\label{fn4.1}\footnote{The product form of (\ref{eq4.1}) can be derived (§ \ref{sec2.3}) by maximizing the entropy $S[\hat \cD (0)]$ and accounting for the constraints ${\rm tr}_A\hat \cD (0) = \hat r(0)$, ${\rm tr}_\rmSB\hat \cD (0) = \hat R_\rmM (0)$, ${\rm tr}_\rmSM\hat \cD (0) = \hat R_\rmB (0)$ through Lagrange multipliers.}

\BEQ \label{eq4.1}
\hat \cD (0) = \hat r(0) \otimes \hat \cR(0) , \quad \hat \cR(0)=\hat R_\rmM (0) \otimes \hat R_\rmB (0).            
\EEQ

The subsequent {\it evolution} of S+A in the Hilbert space $\cH$ is governed by the Liouville--von Neumann equation (\ref{eq2.2}) for the isolated system S+A, with
\BEQ \label{eq4.2}
\hat H   =\hat H_A + \sum_i \hat \pi_i \otimes \hat h_i ,    \qquad  \hat H_A=\hat H_\rmM  + \hat H_\rmB  + \hat H_\rmMB .      
\EEQ
This dynamical equation can be simplified by relying on the {\it conservation laws} which characterize ideal measurements  (§ \ref{sec3.2.2}).
The commutation relations $[\hat H  , \hat \pi_i]=0$ allow us to split it into$^{\ref{fn14}}$ 
\BEQ \label{eq4.3}
&& \rmi\hbar  \frac{\d\,\hat \pi_i\hat \cD(t)\hat \pi_j}{\d t} 
          = [\hat H _A, \hat \pi_i\hat \cD (t)\hat \pi_j] 
+ \hat h_i\times\hat \pi_i\hat \cD (t)\hat \pi_j  - \hat \pi_i\hat \cD (t)\hat \pi_j\times\hat h_j .	
\EEQ
The solution of  (\ref{eq4.3}) with the initial condition (\ref{eq4.1}) has the form 
\BEQ \label{eq4.4}
\hat \pi_i\hat \cD (t)\hat \pi_j = \hat \pi_i\hat r(0)\hat \pi_j \otimes \hat \cR_{ij}(t) ,               \EEQ
where the operators $\hat R_{ij}(t)$ in the space $\cH_A$ are determined by 
\BEQ \label{eq4.5}
\rmi\hbar \frac{ \d\hat \cR_{ij}(t)}{\d t} = [\hat H_A, \hat \cR_{ij}(t)] + \hat h_i \hat \cR_{ij}(t) - \hat \cR_{ij}(t)  \hat h_j ,   
\EEQ
with the initial conditions
\BEQ \label{eq4.6}
\hat \cR_{ij}(0)= \hat R_\rmM (0)\otimes\hat R_\rmB (0).                                      
\EEQ
In the special case where $\hat r(0) \equiv  \hat \pi_i \hat r(0)\hat \pi_i$,  there is a single relevant block  $\hat \cR_{ii}$ and we recover (\ref{eq3.7}) for it.

Owing to the conservation laws which express ideality (§ \ref{sec3.2.2}), the equations of motion (\ref{eq4.4})--(\ref{eq4.6}), associated with the separate sectors of S,
involve {\it only the apparatus} A. The tested system S contributes only through the sources $\hat h_i$ in the Hilbert space $\cH_\rmM $. Dynamics thus exhibit the major role played by the apparatus in an ideal measurement process.

\renewcommand{\thesubsection}{\thesection\arabic{subsection}}
\renewcommand{\thesubsubsection}{\thesubsection.\arabic{subsubsection}}
\subsection{Statistical mechanics approach}
\label{sec4.2}

	Explicit solutions of the equations (\ref{eq4.5}) may be worked out for sufficiently simple models in courses of statistical mechanics, either as formal lectures or as illustrative exercises. An extensive review of such models is presented in  \cite{allahverdyan2013understanding}, which also displays a detailed study of the Curie--Weiss model, including a discussion of the validity of various needed approximations. 

We complement here such approaches with a few comments of tutorial interest. In Appendix A, we discuss the role of the bath and recall how, 
for a weak coupling $\hat H _\rmMB$,  in which case standard approximations of quantum statistical mechanics can be used, 
 the equations (\ref{eq4.5}) for A =M+B can further be reduced to equations for the {\it sole measuring device} M. 

	According to (\ref{eq4.5}), {\it two independent types of evolution} of S+A {\it coexist} during an ideal measurement process, with two {\it different relaxation times}. Each {\it diagonal} block $\hat \cR_{ii}(t)$ obeys the same equation (\ref{eq3.7}) as if $\hat s$ took initially the value $s_i$, and it will thus end up as $\hat \cR_{ii}(\tf)=\hat \cR_i$ characterized by (\ref{eq3.4}). Owing to the {\it action} of S {\it on} M (implemented through a sufficiently large coupling factor  $\hat h_i$  in the interaction $\hat H_\rmSM$), 
this process establishes through (\ref{eq4.4}) a {\it correspondence} within $\hat \cD(\tf)$ between the final macroscopic {\it indication} $A_i$ of the pointer 
and the {\it eigenvalue} $s_i$ of $\hat s$. We thus term it ``{\it registration}'' (\cite{allahverdyan2013understanding}, Sec. \ref{sec7}). 
	
Each {\it off-diagonal block} $\hat \cR_{ij}(t)$ ($i\neq j$) of $\hat \cD(t)$ decays and vanishes  well  before $t=\tf$. We term this disappearance ``{\it truncation}'' 
of the state $\hat \cD(t)$ of S+A. It arises from the 
non-Hamiltonian terms of (\ref{eq4.5}) for $i\neq  j$, 
which come out because the action $\hat h_i$ of S on M is not the same for the eigenvalues $s_i$ and $s_j$.
In return, the action of M on S results in the {\it ``reduction''} of the marginal state $\hat r(t)={\rm tr}_A\hat \cD(t)={\rm tr}_\rmM  D(t)$, according to $\hat \pi_i \hat r(\tf)\hat \pi_j =0$ for $i\neq j$. Depending on the model, truncation may be achieved by different mechanisms
(dephasing by the many degrees of freedom of M, or decoherence due to the coupling with B)
 of pedagogical interest, involving or not the bath (\cite{allahverdyan2013understanding}, Secs. \ref{sec5} and \ref{sec6}). 

As we shall see in § \ref{sec5}, this disappearance of the off-diagonal blocks, as usual in decoherence, is not sufficient to solve the measurement problem.
 Indeed, we cannot yet ascertain (§ \ref{sec5.2}) that each diagonal block is physically meaningful in terms of the outcomes of the pointer.

Simple illustrative exercises, where the bath is disregarded, are worked out in Appendix \ref{secB}. It is instructive to discover how the interaction of S with M, in a dynamical process designed to test the observable $\hat s$, {\it destroys the $q$-information} pertaining to the observables of S {\it incompatible with} $\hat s$, initially embedded in the off-diagonal blocks of $\hat r(0)$. 
In appendix  \ref{secB.3},  we also acknowledge how this irreversibility arises through a {\it cascade of $q$-correlations.}

In Appendix \ref{secC}, we analyze the features which allow us to overcome the {\it irreversibility paradox} in the dynamical study of measurement processes. 
It turns out that the very understanding of measurement theory requires the use of {\it approximations} which, however, become fully justified
in the limit  of large M and B (as in the paradox of irreversibility in statistical mechanics for a macroscopic piece of material).

\renewcommand{\thesubsection}{\thesection\arabic{subsection}}
\renewcommand{\thesubsubsection}{\thesubsection.\arabic{subsubsection}}
\subsection{Thermodynamic approach}
\label{sec4.3}

 The task is to show that the quantum dynamics brings us from the metastable to the stable states. 
This has been  achieved in great detail in ref.  \cite{allahverdyan2013understanding}. 
We bypass here the solution of the above dynamical equations, and directly construct the final state $\hat D(\tf)={\rm tr}_\rmB \hat \cD(\tf)$ of S+M 
by relying on an argument which students should readily accept$^{\ref{footn4}}$. As in macroscopic thermodynamics, we admit that S+M, 
which is an open system weakly coupled to an infinitely large bath B, {\it relaxes towards thermodynamic equilibrium} at the temperature $T$ of B. 
This property can be checked from  (\ref{eq4.5}) on models where the interactions are sufficiently intricate to irreversibly raise the entropy 
$S[\hat D(t)]$ of S+M  up to the largest value compatible with the conservation laws. 
The density operator provided by the {\it maximum entropy} criterion (§ \ref{sec2.3}) then appears not only as the {\it least biased} one 
 \cite{elsasser1937quantum,jaynes1957information12,balian2007microphysics12}, 
but also describes the {\it stable} equilibrium reached after relaxation  \cite{allahverdyan2013understanding}.

\renewcommand{\thesubsection}{\thesection\arabic{subsection}}
\renewcommand{\thesubsubsection}{\thesubsection.\arabic{subsubsection}}
\subsubsection{Equilibrium states of S+M}

\label{sec4.3.1}

As a preliminary step, which may constitute an exercise of statistical mechanics,  we determine the general form of the thermodynamic equilibrium states $\hat D_\eq$ of the system S+M weakly coupled to a thermal bath B. As indicated at the end of § \ref{sec2.3}, we first gather the {\it constraints} on $\hat D$. {\it Normalization} fixes ${\rm Tr}_\rmSM\hat D =1$. The weak interaction of S+M with the bath fixes the $q$-expectation value ${\rm Tr}_\rmSM\hat D(\hat H_\rmM +\hat H_\rmSM)$ of its {\it energy}.  Furthermore, the conserved quantities $\hat x$ defined by (\ref{eq3.5}) provide 
{\it constants of the motion}  which result in constraints. These constraints are expressed as ${\rm Tr}_\rmSM\hat D\hat x_\alpha  = \langle \hat x_\alpha  \rangle$  where we denote as $\hat x_\alpha$  the elements of a basis of linearly independent operators in the vector space of observables $\hat x$. (For a non-degenerate spectrum of $\hat s$, the $\hat x_\alpha$ 's are the projectors $\hat \pi_i$.).

The resulting thermodynamic equilibrium state $\hat D_\eq$ is determined by looking for the maximum of the von Neumann entropy 
 $S(\hat D) = -\Tr \hat D\ln  \hat D$, the above constraints being taken into account by means of 
 Lagrange multipliers $\gamma$, $\beta$,  and $\lambda_\alpha$, respectively. 
 The stationarity condition (\ref{eq4.2}) then provides for $\hat D_\eq$ the {\it generalized canonical} form\footnote{The inequality $\Tr \hat D (\ln  \hat D -\ln  \hat D_\eq) >  0$, valid for any $\hat D$  $(\neq \hat D_\eq)$ satisfying the same constraints as $\hat D_\eq$, shows that this stationary value is a {\it maximum}. The result (\ref{eq4.7}) may alternatively be found by looking for the maximum of the Massieu thermodynamic potential $S(\hat D) - \beta  {\rm Tr}_\rmSM \hat D (\hat H_\rmM +\hat H_\rmSM)$ for given ${\rm tr}_\rmSM \hat D \hat x_\alpha$. Note that the constraints are overabundant, since they include the unit operator which is a linear combination of the $\hat x_\alpha$'s (for instance, for a spin $\frac{1}{2}$: $\hat \pi _\uparrow +\hat \pi_\downarrow =\hat I_2$).}
\BEQ \label{eq4.7}
\ln\hat D_\eq  + \gamma \hat I_\rmSM + \beta (\hat H_\rmM +\hat H_\rmSM) + \sum_\alpha  \lambda_\alpha \hat x_\alpha  = 0,        
\EEQ
where $\hat H_\rmSM = \sum_i\hat \pi_i \otimes\hat h_i$ and where  $\hat I_{\rmSM}$ is the identity operator in $\cH_\rmS\otimes\cH_\rmM$. 

The explicit derivation of $\hat D_\eq$ from (\ref{eq4.7}) may be taught as an algebraic exercise. The operator $\hat X\equiv  \beta (\hat H_\rmM +\hat H_\rmSM) + \gamma \hat I_\rmSM + \sum_\alpha  \lambda_\alpha \hat x_\alpha$  satisfies $[\hat \pi_i,\hat X]=0$ or $\hat \pi_i\hat X\hat \pi_j=0$ for $i\neq  j$, which entails $\hat \pi_i\hat D_\eq\hat \pi_j=0$ for $i\neq  j$. The equality $\hat \pi_i \ln\hat D_\eq\hat \pi_i = -\hat \pi_i \hat X$ then provides 
   $$ \hat \pi_i\hat D_\eq\hat \pi_i =\hat \pi_i \exp (-\hat \pi_i\hat X) = \hat \pi_i \exp [-\hat Y],\qquad
   \hat Y= \beta \hat \pi_i\otimes(\hat H_\rmM +\hat h_i)
     +\gamma \hat \pi_i\otimes\hat I_\rmM  + \sum_\alpha  \lambda_\alpha \hat \pi_i \hat x_\alpha \otimes\hat I_\rmM ,   $$ 
from which we obtain
\BEQ
\hspace{-7mm}	\hat D_\eq = \sum_i \exp\left (-\sum_\alpha  \lambda_\alpha \hat \pi_i \hat x_\alpha-\gamma \right ) \otimes
\exp\left [-\beta  (\hat H_\rmM +\hat h_i) \right ].      
	\EEQ
In the space $\cH_\rmM $,  for each $i$, the last factor provides (within normalization) the equilibrium state $Z\times\hatRih$ of Eq. (\ref{eq3.3}) 
that M alone would reach at the bath temperature $T=1/ \beta$  if its Hamiltonian were $\hat H_\rmM  + \hat h_i$. In the space $\cH_\rmS $, we reparametrize the first 
factor in terms of a scalar $p_i$ and a normalized matrix $\hat r_i$  as
\BEQ
Z\exp\left( -\sum_\alpha  \lambda_\alpha \hat \pi_i \hat x_\alpha-\gamma \right) \equiv  p_i \hatri,         
\EEQ
where $\hatri$ satisfies
\BEQ \label{eq4.10}
 \hatri \equiv  \hat \pi_i \hatri \hat \pi_i ,    \quad    {\rm Tr}_\rmS \hatri =1.                
\EEQ
Normalization of $\hat D_\eq$ implies $\sum_i p_i = 1$.
 
Altogether, the sought thermodynamic equilibrium states of S+M are found as a weighted sum
\BEQ \label{eq4.11}
\hat D_\eq = \sum_i p_i \hatri  \otimes \hatRih  , 				
\EEQ
where $\hatri$ behaves as a density operator confined in the subspace of $\cH_\rmS $ spanned by $\hat \pi_i$. 
The $\hat r_i$'s and the $p_i$'s should be determined from data pertaining to S.

\renewcommand{\thesubsection}{\thesection\arabic{subsection}}
\renewcommand{\thesubsubsection}{\thesubsection.\arabic{subsubsection}}
\subsubsection{The final state}

\label{sec4.3.2}

Admitting that the final state $\hat D(\tf)$ of S+M has the thermodynamic equilibrium form (\ref{eq4.11}),
we determine its parameters  $p_i$ and $\hatri$ by relying on the existence (§ \ref{sec3.2.2}) of the constants of the motion 
${\rm Tr}_\rmSM\hat D(t)\hat x_\alpha  = \langle \hat x_\alpha  \rangle$.
 During the dynamical process, they keep some memory of the initial state $\hat D(0) = \hat r(0) \otimes \hat R_\rmM (0)$. 
 As the conserved observables $\hat x$ have the form  (\ref{eq3.5}), the identity of ${\rm Tr}_\rmSM\hat D(\tf)\hat x$ and ${\rm Tr}_\rmSM\hat D(0)\hat x$ yields
\BEQ \label{eq4.12}
{\rm Tr}_\rmS  \sum_j p_j \hatri \sum_i \hat \pi_i \hat y \hat \pi_i = {\rm Tr}_\rmS  \hat r(0) \sum_ i \hat \pi_i \hat y \hat \pi_i       
\EEQ
for any operator $\hat y$ of $H_\rmS $. Using (\ref{eq4.10}) and the arbitrariness of $\hat y$, we find\footnote{Eq. (\ref{eq4.13}) results from the conservation of $\hat\pi_i$.  Eq. (\ref{eq4.14}) reduces to $\hat r_i = \hat \pi_i$ for a non-degenerate eigenvalue $s_i$ of $\hat s$.}
\BEQ \label{eq4.13}
p_i ={\rm Tr}_\rmS  \hat r(0) \hat \pi_i ,   				
\EEQ
\BEQ \label{eq4.14}
\hat r_i =\frac{ \hat \pi_i \hat r(0) \hat \pi_i  }{ {\rm Tr}_\rmS  \hat r(0) \hat \pi_i } ,	
\EEQ
 
Thus, as long as $\hat H_\rmSM$ is switched on, S+M  reaches the equilibrium state $\hat D_\eq(t)$ of Eqs. (\ref{eq4.11}), (\ref{eq4.13}), (\ref{eq4.14}). We assumed in § \ref{sec3.2.3} that, a little before the end of the process, the interaction $\hat H_\rmSM$ is switched off.
Then, if the equilibrium states $\hat R_i$ of M are well separated, the dynamical equation for the state $\hat R(t)=\tr_\rmS \hat D (t)$ will lead $\hat R(t)$ from $\hat R_i^h$ given by (\ref{eq3.3}) to $\hat R_i$, defined mathematically in § \ref{sec3.2.1} as the limit for $\hat h_i \to0$ of $\hat R_i^h$.

Altogether, the set $\cE$  of compound systems S+M initially prepared in the state $\hat D(0) = \hat r(0) \otimes \hat R_\rmM (0)$ is represented, 
at the {\it final time} $\tf$ of the process, by the density operator
\BEQ \label{eq4.15}
\hat D (\tf) =\sum_i p_i \hat D_i ,   \quad     \hat D_i \equiv  \hat r_i \otimes\hat R_i .          
\EEQ
The weights $p_i$ and the operators $\hatri$ keep partly track of the initial state $\hat r(0)$ of the tested system S through (\ref{eq4.13}), (\ref{eq4.14}), while $\hat R_i$ is characterized by the macroscopic pointer value (\ref{eq3.4}). The bath is so large that it is practically unaffected in spite of its perturbation through $\hat H_\rmMB$ 
and its change of free energy, so that the final state of S+M+B issued from (\ref{eq4.1}) is 
\BEQ \label{eq4.16}
\hat \cD(\tf) =\sum_i p_i \hat r_i \otimes\hat R_i\otimes\hat R_\rmB (0) .           
\EEQ 
 
 This result encompasses everything we can tell within the quantum formalism about S+A at the time $\tf$, knowing the initial state and the dynamics. We cannot go further and give a physical meaning to the separate elements of (\ref{eq4.16}) as long as the pointer indications are not read, as will be explained below (§  \ref{sec5.2}).
  However, this expression is the best possible quantum mechanical description {\it for the full set} $\cE$ of systems S+M+B at the time $\tf$. As regards the system S after separation from A, our knowledge of its final density operator $\hat r(\tf)=\tr_\rmA \hat \cD(\tf) =\sum_i p_i\hat r_i$ allows us to make predictions about further experiments performed on S,  in case we do not observe the pointer and pick up at random the outcomes of the process.

\renewcommand{\thesection}{\arabic{section}}
\section{The measurement problem revisited}
\renewcommand{\thesection}{\arabic{section}.}

\label{sec5}
According to the definition of a quantum state (§ \ref{sec2.1.3}), the final density operator $\hat\cD(\tf)$ of S+A  describes {\it formally} the outcome of the 
{\it full set} $\cE$ of runs, for an initial state $\hat\cD(0)$ of S+A. {\it Physical interpretation} of $\hat\cD(\tf)$ requires,  as already stressed and recalled below, 
to extract  from this operator properties of {\it individual runs}. This is impeded by a purely quantum difficulty (§ \ref{sec5.2}), which we must overcome (§ \ref{sec5.3}).

\renewcommand{\thesubsection}{\thesection\arabic{subsection}}
\renewcommand{\thesubsubsection}{\thesubsection.\arabic{subsubsection}}
\subsection{Measurements and individual runs}
\label{sec5.1}

Our purpose is to explain, by studying {\it ideal} measurements as a model, the following features of {\it real} experiments.

\begin{itemize}

\item	Information about a measurement are generally obtained in the form of data $A_i$ carried by the pointer of the apparatus. We focus on discrete data, defined within some resolution. In the ideal measurement model, these data $A_i$ are associated, according to Eq. (3.4), with the specific states $\hat R_i$ of M. A good resolution,
such that their interval is large compared to the $q$-fluctuation of $\hat A$, requires a proper apparatus with a macroscopic pointer.

\item	Observation of the data $A_i$ requires a safe registration by the apparatus during a sufficiently large delay. The   ideal measurement model  defined in  § \ref{sec3} 
readily satisfies this property. If reading or registration take place while S and A are still coupled,\footnote{For instance, in ref.  \cite{donker2018quantum}, the reading-off is done as late as possible before decoupling.} it is ensured by the stability of the equilibrium state $\hat R_i^h$ defined by (3.3). After separation of S from A, the state $\hat R_i$ is not rigorously 
stable,$^{\ref{footn3.1}}$   but its lifetime is extremely long, allowing ample time for reading off or registering the outcome.

\item
	Among the large set of runs of a measurement$^{\ref{fn11}}$, 
	the outcome of each one is observed individually, yielding a statistic of the possible values of $A_i$, and hence of the possible values of $s_i$.

\item	For a non-demolishing measurement, the selection of runs with a given outcome $A_i$ constitutes a preparation of S in a new, collapsed state.
\end{itemize}

The last two {\it experimental} features are based on the occurrence of a definite answer, characterized by the final value $A_i$ of the pointer variable 
for each individual process.  In spite of the banality of this property, 
we have to explain it {\it theoretically}, by relying on our only knowledge, the initial state of S+A, and on quantum dynamics. 
Then, quantum theory provides nothing more than the result (\ref{eq4.16}) for $\hat \cD(\tf)$, 
which formally describes the outcome of the full set of runs but does not refer to such different {\it exclusive events} occurring in individual runs.	

No problem arises in the special case when the initial state is pure, $\hat r(0)=\hat\pi_i \hat r(0)\hat\pi_i=\hat r_i$,  {lying
 in the eigenspace associated with the single eigenvalue $s_i$. 
Then, the theoretical approach of § \ref{sec3.2.3} yields as final density operator $\hat\cD(\tf) =\hat\cD_i$. 
Consequently, as physically expected, the marginal states of M and S are found as $\hat R_i$ and $\hat r_i$, respectively, implying that each individual run provides the pointer value $A_i$. 
Thus the state $\hat r_i$ of S is left unchanged for any subset of runs. 

However, for an arbitrary initial state $\hat r(0)$ of S, we cannot a priori infer from $\hat \cD(\tf)$ that each run yields a definite indication $A_i$ of the pointer. Nor can we split the expression (\ref{eq4.16}) of $\hat \cD(\tf)$ into meaningful terms in spite of its suggestive form\footnote{In the Curie-Weiss model, the initial state $\hat r(0)=\hat\pi_\uparrow$  of S provides the final state $\hat \cD(\tf) =\hat \cD_\Uparrow= \hat\pi_\uparrow \otimes \hat R_\Uparrow\otimes \hat R_B(0)$ of S+A for the full set, and also for any subset of runs, hence for individual runs. However, if the tested spin lies initially in the $x$-direction, we get $\hat \cD(\tf) =\half(\hat \cD_\Uparrow+\hat \cD_\Downarrow)$ as final state for the full set of runs. 
Then, as indicated in$^{\ref{footn3.1}}$, 
all subsets are described by final states of the form $\lambda\hat \cD_\Uparrow+(1-\lambda)\hat \cD_\Downarrow)$ but the values of $\lambda$ remain unknown. We cannot ascertain the existence of two complementary subsets with $\lambda=1$ and $\lambda =0$. Nothing prevents us even, within the sole quantum formalism, from assigning the same density operator $\half(\hat \cD_\Uparrow+\hat \cD_\Downarrow)$ to any subset; then, the value of the pointer would be undetermined for any individual run.}. As will be shown in § \ref{sec5.2}, the {\it quantum principles alone} do not suffice to achieve these goals. It is impossible to derive from the sole knowledge of $\hat\cD(\tf)$ a physical interpretation for its elements $p_i$ and $\hat r_i$. This measurement problem is a specifically quantum difficulty issued from Schr\"odinger's ambiguity.

\renewcommand{\thesubsection}{\thesection\arabic{subsection}}
\renewcommand{\thesubsubsection}{\thesubsection.\arabic{subsubsection}}
\subsection{Schr\"odinger's ambiguity}
\label{sec5.2}

The above impossibility of directly giving a physical meaning to the separate terms of (\ref{eq4.16}) arises from Schr\"odinger's  {\it ambiguity of decompositions}  
of a non-pure density operator  \cite{park1968nature,allahverdyan2013understanding} (also known as the impossibility of an ignorance-interpretation of mixtures).
We saw in § \ref{sec2.2.3} that the merger of two sets $\cE_1$ and $\cE_2$ of systems represented by density operators $\hat \cD_1$ and 
$\hat \cD_2$ produces a set $\cE$ described by a linear combination of $\hat \cD_1$ and $\hat \cD_2$ (Eq. (\ref{eq2.3})).
Given conversely a density operator $\hat \cD$ describing  a physical set $\cE$,  without any further information, consider a sum into which it can be decomposed,
	\BEQ \label{eq5.1}
\hat \cD = \sum_k \lambda_k \hat \cD_k ,	
\EEQ
where the operators $\hat \cD_k$ have the mathematical properties of density operators (§ \ref{sec2.1.2}) and where the weights $\lambda_k$ are positive and normalized.  (The final state (\ref{eq4.16}) of S+A is such a sum.)  The form of this expression suggests that the state $\hat \cD_k$  might represent some physical subset $\cE_k$ of $\cE$.  Nevertheless, $\hat \cD$ has also many other decompositions
\BEQ  \label{eq5.2}
\hat \cD = \sum_j \mu_j \hat \cD'_j ,			
\EEQ
which suggest that the state $\hat\cD'_j$ might represent some other subset $\cE_j'$ of $\cE$. Then, we will show that one can always
 find pairs (\ref{eq5.1}), (\ref{eq5.2}) of {\it mathematical} decompositions such that a {\it contradiction} would arise if there existed {\it physical}
subsets $\cE_k$ and $\cE_j'$ represented by $\hat \cD_k$ and $\hat \cD'_j$.
Hence, if no other information than $\hat \cD$ is available, the elements of its mathematical decompositions {\it cannot be interpreted as density operators associated with physical subsets of}  $\cE$. A general proof is given in Appendix \ref{secD}, suited to be taught as exercises. 

Let us demonstrate this point on a simple illustrative example. Consider an set $\cE$ of unpolarized spins $\half$  (or of fully random $q$-bits), described by the quantum state $\hat \cD=\half \hat I$. The operator $\hat \cD=\half \hat I$ can be decomposed into an incoherent sum $\hat \cD = \half  (| +z\rangle \langle +z|  + | -z\rangle \langle -z| )$ of two projectors associated with the orientations $+z$ and $-z$, respectively. Alternatively, $\hat \cD=\half \hat I$ can be decomposed into a sum $\hat \cD = \half  (| +x\rangle \langle +x|  + | -x\rangle \langle -x| )$ involving the orientations $+x$ and $-x$. If these mathematical decompositions of $\hat \cD$ were both physically meaningful, we would infer that $\cE$ might be split into two subsets $\cE_{+z}$ and $\cE_{-z}$ gathering individual spins polarized along $+z$ and $-z$, respectively -- but that it might also be split into two subsets $\cE_{+x}$ and $\cE_{-x}$ gathering spins polarized along $+x$ and $-x$. Then, there would exist a subset $\cE_{+z} \cap \cE_{+x}$ of $\cE$, including a quarter of the spins, which would gather spins simultaneously polarized along $+z$ and $+x$, which is absurd. 

The ambiguity of mathematical decompositions of $\hat \cD$, which forbids us to give them a physical meaning, occurs for any type of preparation, due to issues such as
\begin{itemize}
\item (a) {\it Lack of control} of some variables. 

\item (b) {\it Stochastic evolution}, such as depolarization.

\item  (c) {\it Merger}. 
 Consider two sets $\cE_1^{(+z)}$ and $\cE_1^{(-z)}$  of systems made of spins $\half$ polarized in the pure states$| +z\rangle$  and $| \! -z\rangle$, respectively. 
 Somewhat artificially, we extract from them an equal number of individual systems, and gather them without keeping track of the ensemble they came from
  and forgetting the axis $\pm z$ along which they had been polarized. 
 The set $\cE_1$ resulting from this merger is described, according to (2.3), by the state
 $\hat \cD_1 = \half  (| +z\rangle \langle +z|  + | -z\rangle \langle -z| )=\half\hat I$. Nothing else than $\hat \cD_1=\half\hat I$ is known about this set $\cE_1$. 
 A similar merger, starting from sets $\cE_2^{(+x)}$ and $\cE_2^{(-x)}$  such that the spins are polarized along $\pm x$ instead of $\pm z$, produces a new 
 set $\cE_2$ which is described by the state  $\hat \cD_2 = \half  (| +x\rangle \langle +x|  + |-x\rangle \langle -x| )=\half\hat I$. 
 However, as $\hat \cD_1 = \hat \cD_2 = \half \hat I$, no experiment can distinguish the sets $\cE_1$ and $\cE_2$. 
 This impossibility of retrieving, in the quantum formalism, the physical difference between $\cE_1$ and $\cE_2$ arises from the {\it loss of the original information} 
 about the direction of the polarization of each individual spin in the construction of these sets.

\item (d) {\it Partial trace}. The difficulty is flagrant for a further set $\cE_3$ of unpolarized spins prepared by use of an {\it ancilla},
a second system.
 After having produced spin pairs in a singlet state, one extracts the first spin from each pair, disregarding the ancillary second spin (traced out as in § \ref{sec2.2.2}). The resulting set $\cE_3$ is again described by the state $\hat D_3 = \half \hat I$, indistinguishable from $\hat \cD_1$ and $\hat \cD_2$. Its preparation is isotropic, so that its discrete decompositions, which involves privileged directions, cannot be physically relevant. 

\item (e) {\it Approximate dynamics}.  For a macroscopic system that relaxes to equilibrium, approximate methods of statistical mechanics (Appendix C) 
produce a final state which is mixed, therefore ambiguous. In measurements, the ambiguity of $\hat \cD(\tf)$ is the result of such dynamics. 

\end{itemize}

This ambiguity, a peculiarity of quantum descriptions  (for non-pure states), has no equivalent for ordinary probabilities, which characterize the relative number of individual 
objects  possessing some property  \cite{park1968nature}. There, each splitting of the probability distribution directly reflects a splitting of the considered physical set $\cE$ 
(which is then an ordinary statistical ensemble) 
into subsets (see end of Appendix \ref{secD}).  In quantum mechanics, however,  the matrix nature of ``states'', which arises from the non-commutation 
of observables, precludes the existence of a physical subset of $\cE$ that would be described by a quantum formal entity such as $\hat \cD_k$ in (\ref{eq5.1}). 
As already stressed, a density operator is an abstract object that only refers to an {\it overall set} of systems, {\it not to a sample space} made of individual systems; it {\it cannot be split} into physically significant parts if nothing else is known.

\renewcommand{\thesubsection}{\thesection\arabic{subsection}}
\renewcommand{\thesubsubsection}{\thesubsection.\arabic{subsubsection}}
\subsection{A postulate concerning exclusive events} 
\label{sec5.3}

The measurement problem arises from the discrepancy between the expected physical features of individual ideal measurements (§ \ref{sec5.1}) and the abstract quantum description 
of the large set $\cE$ of systems S+M  by the density matrix $\hat D=\tr_\rmB \hat \cD$. Knowing only the density operator $\hat D(\tf)$ issued from $\hat D(0)$ and expressed 
by (\ref{eq4.15}), we are unable to extract information about individual runs.
Indeed, due to the above ambiguity (§ \ref{sec5.2}), we cannot directly assert that the ingredients $\hat D_i$ of its decomposition $\hat D(\tf)=\sum_i p_i \hat D_i$ describe subsets, although this looks tempting -- and will turn out to be correct. Students should be warned that more information than encoded in $\hat D(\tf)$ is needed to overcome its ambiguity  \cite{park1968nature}. 

To deal with this difficulty, we acknowledge the existence of systems ``M'' having the following property $\cP$ inferred from macroscopic experience:

\begin{itemize}

\item   $\cP$:  
A physical system ``M'' is an object which possesses several possible equilibrium states $i$ characterized by the value $A_i$ of some quantity, and which, 
under the considered conditions of preparation, reaches either one or another among them at the issue of each individual run of the experiment. 
We postulate that, not only a large set of such systems M, but also each of its large 
subsets distinguished by the value $A_i$, can be described in quantum theory  by a density operator.
\end{itemize}


	Examples of systems M are provided in statistical physics by materials with broken invariance, the variable $A_i$ being the order parameter. For instance, the Ising magnet of § \ref{sec3.2.1} has the required two equilibrium states $\hat R_\Uparrow$ and $\hat R_\Downarrow$ characterized by well-defined magnetizations $+M_\rmF$ or $-M_\rmF$. The material reaches only one of these two states in individual experiments, in spite of the existence of many other equilibrium states$^{\ref{footn3.1}}$ 
	which are not produced in the considered situation.\footnote{The role of the preparation of M in the postulate $\cP$ may be exemplified by superconductivity. Breaking in Fock space the invariance under transformations $\exp(i\chi\hat N)$, where the particle number operator $\hat N$ behaves as a quantum variable, leads theoretically to the BCS equilibrium states $|\Psi_\phi\rangle$ characterized by an order parameter, the phase $\phi$ of the pairing function $\langle \hat\psi_{{\bf k}\uparrow} \psi_{-{\bf k}\downarrow}\rangle$ (defined within $2\pi$). If the superconducting sample M is part of a much larger system, this breaking of invariance is effective: the system reaches an equilibrium BCS state, an eigenstate of a unitary operator $\exp(i \hat\phi)$,  with a phase $\phi$ taking well-defined values (within $2\pi$) for individual samples. (More precisely, in superconducting devices, the well-defined quantities are the phase differences between different subsystems.) The phase $\phi$ can be identified with the variable $A_i$ of $\cP$. However, if the system is produced with a given particle number $N_0$, pairing is theoretically taken into account by first breaking the invariance in Fock space, then by restoring it through projection on the Hilbert space for $N_0$ particles. The equilibrium state is then proportional to$\int_0^{2\pi}\d\phi\, |\Psi_\phi\rangle\exp(-iN_0\phi)$. The phase cannot be defined for individual runs and the property $\cP$ fails. (A similar restoration of broken invariance is currently used to account for pairing in nuclei.)}

A measuring device should be a physical system M satisfying the above property $\cP$. 
We thus acknowledge the existence of a pointer which gives a {\it well-defined indication} $A_i$ at the issue of each {\it individual process}, 
allowing us to tag the runs by $A_i$ at the time $\tf$. Hence, the set $\cE$ of systems S+M tested by the ideal measurement can be split 
(at the time $\tf$) into subsets $\cE_i$, termed as ``{\it homogeneous}'', which gather all events having yielded the outcome $A_i$. 
Reading or registering these outcomes after achievement of the process requires {\it stability}, which is ensured by a {\it macroscopic} size of the apparatus.


The postulate $\cP$ may be regarded as an implementation of the {\it  Heisenberg cut}, a theoretical feature of ideal quantum measurements. Here, a kind of ``cut'' occurs between § 4 and § 5. On the first side (§ 4.2), we describe mathematically the ensemble $\cE$ of isolated systems S+A, between the times $t=0$ and $\tf$, by a density operator that evolves according to the Liouville--von Neumann equation, in the limit of a large A. On the second side (§ 5.3), a partly classical behavior is exhibited at the final time $\tf$ for the pointer observable $\hat A$, as it takes for each individual run, at the issue of the process, a value $A_i$ well-defined at the macroscopic scale, which can thereafter be read by an observer. This property, foreign to standard quantum mechanics which deals with ensembles, affords the required distinction of the subsets $\cE_i$ of $\cE$.

{\it Physically}, the possible indications $A_i$ are {\it exclusive}, so that the occurrence of some value $A_i$ for each individual run appears as a probabilistic event. 
 {\it Ordinary probabilities}  in the sense of relative frequencies come out by counting those which yield each possible result. Thus, as regards the outcomes $A_i$, the set of runs constitute a {\it statistical ensemble} (§ \ref{sec2.1.2}). However, {\it theoretically} we have stressed that such physical features are foreign to the abstract 
 quantum formalism of § \ref{sec2}, from which they cannot be deduced since quantum states refer only to sets. We will match this formalism with ordinary probability theory, which refers to {\it individual systems}, by incorporating the postulate $\cP$ into measurement theory.

To this aim,  we note that, owing to the last part of this postulate,  each homogeneous subset  $\cE_i$ of systems S+M can be described at the time $\tf$ by some density operator 
$\hat\Delta_i$. It now remains to determine $\hat \Delta_i$ from the available data.

\renewcommand{\thesubsection}{\thesection\arabic{subsection}}
\renewcommand{\thesubsubsection}{\thesubsection.\arabic{subsubsection}}
\subsection{Density operators of subsets}
\label{sec5.4}

A rigorous assignment of a density operator $\hat\Delta_i$  to each homogeneous subset $\cE_i$ of systems S+M after the time $\tf$ will rely on two pieces of information (§ \ref{sec2.3}).

On the one hand, after the above property $\cP$ has allowed us to {\it distinguish} the subsets $\cE_i$ within $\cE$, their merger must reconstruct the whole set $\cE$, 
according to (\ref{eq2.3}). We know that $\cE$ is represented at the time $\tf$ by the density operator (\ref{eq4.15}). Matching at $\tf$ the quantum descriptions of the sets $\cE_i$ and $\cE$ yields a {\it constraint on the operators} $\hat\Delta_i$,
\BEQ  \label{eq5.3}
	\sum_i \cN_i \hat \Delta_i = \cN \hat D(\tf) = \cN \sum_i p_i \hat r_i \otimes\hat R_i ,  	
\EEQ
where $\cN_i$ (still to be determined) and $\cN$ are the respective numbers of runs of $\cE_i$ and $\cE$. 
Due to the ambiguity of the decompositions of $\hat D(\tf)$, 
we are not yet allowed to infer from (5.3) that $ \cN_i \hat \Delta_i =\cN p_i\hat r_i\otimes\hat R_i$.

On the other hand, each subset $\cE_i$ is characterized by the macroscopic value $A_i$ of the pointer (§ \ref{sec3.2.1}). We wish to express this property 
as a mathematical constraint on $\hat \Delta_i$. In the Hilbert space $\cH_\rmM $, we extract from the spectrum of $\hat A$ the eigenvalues lying between $A_i - \delta$  and $A_i + \delta$, where $\delta$  is much larger than the $q$-fluctuation of $\hat A$ in $\hat R_i$ and much smaller than the separations between the $A_i$'s. We denote as $\hat \Pi_i$ the projector\footnote{The projectors 
$\hat \pi_i$, associated with $s_i$ in the Hilbert space $\cH_\rmS$, and $\hat \Pi_i$, associated with $A_i$ in the space $\cH_\rmM$\!, are in one-to-one correspondence but have different properties. The dimension of $\hat \Pi_i$ is large as an exponential of the degrees of freedom of M,
of order $ \exp S(R_i)$. While $\sum_i \hat \pi_i = \hat I_\rmS $, the sum $\sum_i \hat \Pi_i$ spans only a subspace of $\cH_\rmM$\!. While $\sum_i s_i \hat \pi_i = \hat s$, the sum $\sum_i A_i \hat \Pi_i$ does not encompass the full observable $\hat A$. While in an ideal measurement each $\hat \pi_i$ is a constant of the motion $([\hat \pi_i,\hat H  ]=0)$, $\hat \Pi_i$ cannot commute with $\hat H$ since the pointer evolves between $t=0$ and $\tf$ (under the effect of $\hat H_\rmMB\!$).\label{fn28fn}}
 on the subspace spanned by the corresponding eigenvectors of $\hat A$. The observable $\hat \Pi_i$ characterizes the occurrence of the macroscopic indication $A_i$. This event occurs with certainty in the subset $\cE_i$, so that (§ \ref{sec2.1.4}) the state $\hat \Delta_i$ is constrained by $\Tr\hat \Delta_i \hat \Pi_i = 1$ in the Hilbert space $\cH_\rmSM$. Owing to the properties of density operators (§ \ref{sec2.1.2}), this property is equivalent to 
\BEQ  \label{eq5.4}
			\hat \Pi_i \hat \Delta_j \hat \Pi_i = \hat \Delta_i \delta_{ij}.      	
\EEQ
We have likewise
\BEQ  \label{eq5.5}
			\hat \Pi_i \hat R_j \hat \Pi_i = \hat R_i \delta_{ij}.      		  
\EEQ
Multiplying then (\ref{eq5.3}) on the left and on the right side by $\hat I_\rmS \otimes\hat \Pi_i$ and making use of (\ref{eq5.4}) and (\ref{eq5.5}), we find
\BEQ \label{eq5.6}
\cN_i \hat \Delta_i = \cN p_i \hat r_i \otimes\hat R_i .                      
\EEQ
This result agrees with a naive guess from (\ref{eq4.15}), but both properties (\ref{eq5.3}) and (\ref{eq5.4}) have been needed to prove it. 

Altogether, we have found the density operator $\hat \Delta_i$ of S+A after the time $\tf$ for the homogeneous subset $\cE_i$ as
\BEQ  \label{eq5.7}
		\hat \Delta_i = \hat D_i = \hat r_i \otimes\hat R_i  ,     		
\EEQ 
and the {\it number $\cN_i$ of runs} of $\cE_i$ as 
\BEQ  \label{eq5.8}
			\frac{\cN_i }{\cN} = p_i .			
\EEQ
Through the expressions (\ref{eq4.13}) and (\ref{eq4.14}) of $p_i$ and $\hatri$, these equations exhibit how the compound system S+M retains, after selection of the pointer indications, a partial memory of the initially prepared state $\hat r(0)$ of S.

\renewcommand{\thesection}{\arabic{section}}
\section{Interpretation and discussion}
\renewcommand{\thesection}{\arabic{section}.}

\label{sec6}

\renewcommand{\thesubsection}{\thesection\arabic{subsection}}
\renewcommand{\thesubsubsection}{\thesubsection.\arabic{subsubsection}}
\subsection{Macroscopic features and consistency}
\label{sec6.1}

Although the quantum formalism does not describe individual systems, their consideration in the property $\cP$ of § \ref{sec5.3} 
has allowed to introduce the subsets $\cE_i$ of $\cE$, which could thereafter be represented by the quantum states $\hat D_i$ of (\ref{eq5.7}). 

The above analysis shows how the postulate $\cP$, issued from  pre-quantum physics, can be conciliated with the quantum formalism -- 
although explaining it would require new ideas. It leads, in the sense of § \ref{sec1.2} (c) below,
to a physical interpretation of some quantum quantities. Indeed, this property introduces within quantum mechanics a classical probabilistic structure 
for the outcomes $A_i$ which constitute a standard sample space of probabilistic events. Thus, the {\it abstract $q$-probability }
\BEQ  \label{eq6.1}
p_i=\Tr_\rmSM\hat D(\tf)\hat \Pi_i 				
\EEQ
is identified through (\ref{eq5.8}) with a {\it relative frequency} $\cN_i /\cN$, and is thus interpreted as the {\it ordinary probability
of occurrence,  within a large set of runs,  of the outcome $A_i$ for } M.
The formal $q$-{\it expectation} $\Tr_\rmSM\hat D(\tf)\hat A =\sum_i \cN_i A_i/\cN$ is interpreted as a {\it physical mean value}.

This interpretation holds {\it only after the time} $\tf$ when thermodynamic equilibrium of S+M has been reached. Note also that {\it further decompositions} 
of the density operator $\hat D_i$ associated with the subset $\cE_i$ would be plagued by Schr\"odinger's ambiguity and remain meaningless  \cite{park1968nature}.

The possibility of reading, registering and selecting the pointer outcomes requires the states $\hat D(\tf)$ and $\hat D_i$ to be {\it stable}, not only as maximum entropy states, but {\it also dynamically} for slight shifts off equilibrium  \cite{allahverdyan2013understanding}$^{\ref{footn3.1}}$. 
This theoretical requirement is also consistent with the fact that a further measurement performed on an element of $\cE_i$ yields the result  $A_i$.

\renewcommand{\thesubsection}{\thesection\arabic{subsection}}
\renewcommand{\thesubsubsection}{\thesubsection.\arabic{subsubsection}}
\subsection{Physical meaning of Born's rule}
\label{sec6.2}

The above interpretation of (\ref{eq6.1}) still refers only to the {\it apparatus}. The one-to-one correspondence between $A_i$ and $s_i$ exhibited in $\hat D_i = \hatri \otimes\hat R_i$,
together with $\Tr_\rmM\hat R_j \hat\Pi_i=\Tr_\rmS \hat r_j \hat \pi_i=\delta_{ij}$,   provides an alternative expression for $p_i$,
\BEQ \label{eq6.2}
		p_i=\Tr_\rmSM\hat D(\tf)\hat \pi_i = \Tr_\rmS \hat r (\tf) \hat \pi_i ,		
		\EEQ
which now refers to the {\it tested system}$^{\ref{fn28fn}}$. 
 Observing the pointer yields {\it indirect} information about S. We can assert that, at the issue of the measurement process, $\hat s$ takes the value $s_i$ for each {\it individual run} tagged by the outcome $A_i$, and that $p_i$ is the relative number of occurrences of the value $s_i$ of the tested observable $\hat s$ of S. {\it For the observable} $\hat s$ (and for functions of $\hat s$), $q$-probabilities, $q$-expectation values, $q$-variances or $q$-correlations (if $\hat s$ is the product of two commuting observables) are thereby identified with ordinary probabilities, mean values, variances or correlations. 
 However, this interpretation which requires a previous interaction with the dedicated macroscopic apparatus) 
  is {\it restricted} to the {\it final equilibrium state} and to the {\it tested observable} $\hat s$.
  
A third expression of $p_i$ results from the dynamical treatment of the measurement process. Conservation of $\hat s$, expressed by $[H, \hat \pi_i] = 0$ 
(while $\hat \Pi_i$ is not conserved), implies that ${\rm Tr}_\rmSM \hat D (t) \hat \pi_i$ does not depend on time, and we get from (\ref{eq6.2})
\BEQ \label{eq6.3}
p_i = \Tr_\rmSM \hat D (0) \hat \pi_i = \Tr_\rmS  \hat r (0) \hat \pi_i  .         			
\EEQ
By tracing out the apparatus, we have recovered the {\it standard form of Born's rule:}
\BEQ \label{eq6.4}
p_i = \Tr_\rmS  \hat r (0) \hat \pi_i ,    \quad   \langle \hat s\rangle  = \sum_i p_i s_i = \Tr_\rmS  \hat r (0) \hat s .  
\EEQ

Students should not be misled by the expressions (\ref{eq6.4}), which involve {\it only} S {\it before interaction with} A, and which may suggest a physical interpretation 
of $p_i$ in terms of the {\it sole initial state} $\hat r (0)$ of S.  But, rather, they are a result of a {\it dynamical conservation law} of $\hat s$, the presence of which is 
one of the conditions for ideality of the measurement. Moreover, the commutation of a state $\hat\rho(0)$ with the set of projectors $\hat\pi_i$ associated with the eigenvalues $s_i$ 
of $\hat s$ would be needed to interpret the occurrences of $s_i$ as events of a probabilistic sample space (§ \ref{sec2.1.4}). 
As $\hat r(0)$ {\it does not commute} in general with the observables $\hat\pi_i$, Born's $q$-probability ${\rm Tr}_\rmS \hat r (0)\hat\pi_i$  {\it cannot be regarded, at the time $t = 0$, as a probability} in the sense  $p_i = \cN_i /\cN$. This confirms that we should regard the ``state'' $\hat r(0)$ as an abstract {\it mathematical tool} for making probabilistic predictions about the set of runs of any future experiment.
However, while $p_i$ can be interpreted as a probability of occurrence of the eigenvalue $s_i$ at the {\it final time} $\tf$, 
one should not forget that, although the apparatus does not appear in 
its expression $p_i = \Tr_\rmS \hat r (0) \hat \pi_i$, this formula is an indirect consequence of the ordinary probabilities 
$p_i=\Tr_\rmSM\hat D(\tf)\hat\Pi_i = \Tr_\rmM\hat R(\tf) \hat\Pi_i$ of {\it readings of the macroscopic pointer}. 

In fact, (\ref{eq6.4}) is the result of a {\it double inference}, first from M to S at the final time $\tf$, then from $\tf$ to the initial time $t_i=0$. 
The experimental determination from M of the frequencies $p_i$ at $\tf$ allows a partial retrodiction, yielding abstract $q$-information on $\hat r(0)$ through (\ref{eq6.4}).

\renewcommand{\thesubsection}{\thesection\arabic{subsection}}
\renewcommand{\thesubsubsection}{\thesubsection.\arabic{subsubsection}}
\subsection{Physical meaning of von Neumann's reduction}
\label{sec6.3}

We have termed as {\it truncation} the disappearance of the off-diagonal blocks $\hat R_{ij}(t)$ ($i\neq j$) of the density matrix $\hat\cD(t)$ of S+A (§ 4. 2). 
While being approximate, this disappearance is effective for all practical purposes. 
This was a necessary condition for splitting $\hat\cD(\tf)$ into components describing the subsets $\cE_i$ of $\cE$. 
The resulting change of state of the tested system S is called as usual {\it reduction} or {\it collapse}.

We now focus on the marginal state $\hat r(t)={\rm tr}_\rmM \hat D(t)$ of the tested system S. For the full set $\cE$ of runs, 
the ideal measurement of $\hat s$ leads irreversibly $\hat r(t)$ from $\hat r(0)$ to
\BEQ  \label{eq6.5}
\hat r(\tf) =\tr_A \hat\cD(\tf) =\sum_i \hat \pi_i \hat r(0)\hat \pi_i =\sum_ip_i\hat r_i            .
\EEQ
This state is the one which should be assigned to S at the issue of the measurement {\it if the pointer is not read off} and if, accordingly, systems could not have been selected within $\cE$.}
  During the process, S has been perturbed by its interaction with A, which has eliminated the off-diagonal blocks 
  $\hat \pi_i \hat r(0)\hat\pi_j$ ($ i \neq j$) of $\hat r(0)$.  Their disappearance represents a {\it loss of $q$-information} about observables that 
  do not commute with $\hat s$, which is measured by $S[\hat r(\tf)]-S[\hat r(0)]$. 
This is a {\it price paid to acquire information about} $\hat s$  \cite{allahverdyan2017sub}.\footnote{Measuring the component $\hat s_z$ of a spin destroys the information about 
$\hat s_x$ and $\hat s_y$ embedded in $\hat r(0)$ in such a way that  $\langle\hat s_x\rangle=\langle\hat s_y\rangle=0$ in the state $\hat r (\tf)$, 
a property which can be checked by subsequent ideal measurements of $\hat s_x$  and $\hat s_y$.}
 Note that the entropy $S[ D (t)]$ of the compound system S+A  increases due to its relaxation towards equilibrium.

For each subset $\cE_i$ of $\cE$, the density operator to be assigned to S after the time $\tf$ owing to the reading of the outcome $A_i$ is given by (\ref{eq4.14}) as
\BEQ \label{eq6.6}
\hat r_i =\frac{ \hat \pi_i \hat r (0) \hat \pi_i }{{\rm Tr}_\rmS  \hat r (0) \hat \pi_i }.		
\EEQ
We thus retrieve {\it von Neumann's reduction}, which comes out as the result of three steps: ($i$) {\it joint evolution of S+A} (§ 4); 
($ii$) {\it selection of the runs} tagged at $\tf$ by the value $A_i$ of the pointer (§ \ref{sec5.3}); ($iii$) {\it elimination} of the measurement apparatus M.

The occurrence, for the same object S, of different final states $\hat r(\tf)$ and $\hat r_i$ under the same initial conditions illustrates the interpretation 
of a density operator as a catalogue of knowledge of the {\it q}-probabilities associated with a given set $\cE$ of systems $^{\ref{fn6}}$ 
(§ \ref{sec2.1.3}). Knowing only that S has interacted with an apparatus devoted to the measurement of $\hat s$ 
leads to the assignment of the state $\hat r(\tf)$ to S. Learning moreover that the outcome $A_i$ has been found leads to the assignment of $\hat r_i$ for the accordingly selected set.

The so-called ``{\it collapse}'' leading $\hat r(0)$ to some given $\hatri$ may look paradoxical if one misinterprets it as a physical modification of an {\it individual} system S 
caused by its interaction with M. Of course, we acknowledge that each {\it  individual physical process} leads to one among several different possible outcomes;
 but this property is not theoretically accounted for {\it in the quantum formalism} which only deals with sets.
Before selection of the runs according to the indications of the pointer, the ``state''  $\hat r(t)$ is an {\it indivisible} abstract description of the evolution of $S$ 
for the {\it full set} $\cE$, whereas each  $\hat r_i$ describes {\it after} $\tf$ one among its homogeneous {\it subsets} $\cE_i$, which were embedded beforehand within $\cE$.
The authors think that   there is no sense in trying to explain within  the standard quantum formalism the transition 
from the single state  $\hat r (0)$  of S to one among the states $\hat r_i$ as some quantum process 
that would involve a ``{\it bifurcation}'' in the dynamics of S+A.
In fact, quantum dynamics lead the whole set $\cE$ of systems S+A from the state $\cR(0)$ to $\sum_i p_i \cR_i$, meaning that the sole knowledge 
of the initial state and of the interaction of S with A does not inform us about the state $\hat r_i$ in which a given system S ends up. 
The ``collapse'' from $\hat r(\tf)$ to $\hat r_i$ is {\it not a physical process}. it  results from a selection of the subset $\cE_i$, which, 
grace to the postulate $\cP$, can be described theoretically by $\hat r_i$.

Regarding a density operator as a means of prediction, von Neumann's reduction is understood as an {\it updating of the $q$-information} on S, 
each subset $\cE_i$ being described by $\hat r_i$ which encodes a new set of $q$-probabilities. This selection amounts to a {\it preparation} 
(§ \ref{sec2.3}) of the new quantum state $\hat r_i$ of S. Reading the values $A_i$ of the pointer provides us, on average, with an amount of 
$q$-information on S measured by $S[\hat r(\tf)] - \sum_i p_i S(\hat r_i)$. 

If the eigenvalue $s_i$ is non-degenerate, reduction prepares S in the pure state $\hat r_i = \hat \pi_i$ whatever $\hat r (0)$ (provided $p_i\neq 0$). For a degenerate $s_i$, the reduced state (\ref{eq6.6}) found above, which keeps some memory of the initial state $\hat r (0)$ of S, is the one given by {\it L\"uders's rule}  \cite{luders2006concerning}. However, if this initial state is unknown, and if the result $A_i$ has been read, we should assign to S at the time $\tf$ the most random density operator compatible with the selected eigenvalue $s_i$ of $\hat s$ (§ \ref{sec3.2.2}). The above dynamical process for S+A followed by selection of the outcome $A_i$ then produces an alternative reduced state given by {\it von Neumann's rule} $\hat r_i \propto  \hat \pi_i$.  

This interpretation, based on the consideration of subsets determined by the indications of the apparatus, helps to understand EPR experiments. The system S is a pair of spins S$'$ and S$''$ first entangled in the singlet state $\hat r(0)$ then sent far from each other, in regions X$'$ and X$''$ of space. The observable 
$\hat s'_z = \hat \pi '_\uparrow  - \hat \pi '_\downarrow$  is measured in the region X$'$ with an apparatus M$'$ coupled only to S$'$. 
If Alice observes for some run the outcome $A'_\Uparrow$ of the pointer at the time $\tf$, she is led to assign to S the reduced state 
 $\hat r_\uparrow  = \hat \pi '_\uparrow  \otimes \hat \pi ''_\downarrow$. 
  Hence, owing to the conservation of $\hat s_z'+\hat s_z''$  she can {\it infer} that the spin 
S$''$ located in the distant region X$''$ lies in the marginal state $\hat r''(\tf) =  \hat \pi''_\downarrow$. 
However, although entanglement allows Alice to get instantaneously knowledge at a distance about S$''$, there is no quantum communication at a distance. 
Indeed, before he communicates with Alice, Bob cannot do better than assigning to the spin S$''$an unpolarized marginal state; he can assign to S$''$ the state $\hat r''(\tf)$, 
in view of performing subsequent experiments in the region X$''$, only {\it after Alice has transferred him}, from the region X$'$ towards X$''$, 
{\it the (classical) information} that she acquired by having read the result $A'_\Uparrow$.

\renewcommand{\thesubsection}{\thesection\arabic{subsection}}
\renewcommand{\thesubsubsection}{\thesubsection.\arabic{subsubsection}}
\subsection{``Contextuality'': interpretation of measurement arises via the apparatus}
\label{sec6.4}

Even though the standard properties (\ref{eq6.4}) and (\ref{eq6.6}) of ideal measurements are expressed in terms of S alone, we have exhibited the crucial role played by the apparatus A. To establish them theoretically, we relied on the equations of motion  (\ref{eq4.5}) which involve {\it only} A. To solve the measurement problem, we took advantage of the {\it uniqueness} $\cP$ of the pointer outcomes (§ \ref{sec5.3}), a property of a suitable macroscopic apparatus M.

This enforces the idea (§ \ref{sec2.1.3}) that the quantum ``state'' $\hat r(0)$ of S  can be regarded as a {\it synthesis of predictions} 
 about experiments in which S interacts with {\it any possible} {\it  external} apparatus, or ``context''.\footnote{This idea underlies the CSM approach 
  \cite{auffeves2016contexts,auffeves2020deriving} which appears as a reverse of the present approach. Instead of being postulated as in § \ref{sec2}, the quantum formalism is constructed by starting from the consideration of ``Modalities'' M, 
which are exclusive events as in § \ref{sec5.3}. They involve, together with the ``System'' S, an infinite ``Context'' C, which is an idealization of the pointer of an apparatus and which is in correspondence with a basis of $\cH_\rmS$. A modality gathers a pure state of S and an associated value $A_i$ of a classical variable of C, somewhat like our state $\hat D_i =\hat r_i\otimes \hat R_i$. Each given context generates ordinary probabilities $p_i$. However, the probabilistic structures generated by all possible different contexts are not compatible with an overall probability distribution. Nevertheless, by use of a construction based on Gleason's theorem, that whole set of probabilities can be synthetically encoded in a density matrix for S, probabilities $p_i$ being given by Born's rule.} 
 The full $q$-information embedded in $\hat r(0)$ is latent, and only its part that pertains to the {\it specific observable} $\hat s$ is unveiled 
 in the form of ordinary, usable information, at the issue of the interaction of S with a {\it dedicated apparatus} A. 

The joint measurement of non-commuting observables is an important question, still partly open, both experimentally and theoretically \footnote{Partial information 
about non-commuting observables may, however, simultaneously be gained through non-ideal measurements, 
as shown by solvable models (see  §   8.3 of  \cite{allahverdyan2013understanding} and   \cite{allahverdyan2010simultaneous,perarnau2017simultaneous}).
 Joint measurement of non-commuting observables raises  experimental and theoretical questions. }. 
 There, the above interpretation issued from the consideration of the outcomes of ideal measurements cannot be given, because {\it different apparatuses} acting on 
{\it different samples} 
would be required (§ \ref{sec2.1.4}). However, the situation is simple for {\it two commuting} observables $\hat s'$ and $\hat s''$. The eigen-projectors $\hat \pi’_j$ ($j=1, 2,\cdots$) of $\hat s'$
and $\hat \pi''_k$ ($k=1, 2,\cdots$) of  $\hat s''$ commute with each other, hence can simultaneously take the values $\hat \pi'_j=1$ and $\hat \pi''_k=1$. 
This occurrence defines exclusive events ($j, k$) which constitute a sample space of probability theory. One can then associate with a state $\hat r(0)$ of the  system $\\rmS$
an ordinary probability distribution $P(j, k) =\Tr_{\!  \rmS} \,\hat r (0)\hat \pi'_j \hat \pi''_k$ where $P(j, k)$ is the joint probability of occurrence of the eigenvalues $s’_j$ for $\hat s'$ and $s''_k$ for $\hat s''$.
 Born’s expression $\langle\hat s' \hat s''\rangle   =\Tr_{\!  \rmS} \, \hat r (0) \hat s'\hat s''$ yields ordinary correlations $\langle\hat s' \hat s''\rangle -\langle \hat s'\rangle\langle \hat s''\rangle$. 
 They can physically be determined by means of an ideal measurement performed with a compound apparatus measuring both $\hat s'$ and $\hat s''$.

 
Nevertheless, by putting together ordinary correlations determined by ideal measurements on $\hat r(0)$ that require different experimental settings, one may stumble on results that seem 
{\it physically} (Bell \cite{bell1966problem}) or {\it logically} (GHZ \cite{greenberger1989going}) incompatible. 
Consider for instance,  as in the end of § \ref{sec6.3}, pairs of spins S$'$ and S$''$ previously 
prepared in the entangled singlet state $\hat r(0)$. We focus on the observables $\hat s’_z$, $\hat s’_x$, $\hat s''_u$, $\hat s''_v$ 
referring to the directions $z$, $x$, $u = - (z+x)/\sqrt{2}$, $v= -(z-x)/\sqrt{2}$.  First, Alice measures $\hat s’_z$ and Bob $\hat s''_u$, on each one's member of the same pair. 
Repeating this experiment determines $\langle \hat s’_z\hat s''_u\rangle$, an ordinary correlation since $\hat s’_z$ commutes with $\hat s''_u$. 
A second set of experiments provides $\langle \hat s’_z\hat s''_v\rangle$; they require a different apparatus since $\hat s’_z\hat s''_u$ does not commute with $\hat s’_z\hat s''_v$. 
A third set of experiments provides $\langle \hat s’_x\hat s''_u\rangle$ and a fourth one $\langle \hat s’_x\hat s''_v\rangle$. Students can evaluate as 
an exercise \cite {clauser1969proposed} the quantity 
$C \equiv \langle \hat s’_z\hat s''_u\rangle + \langle \hat s’_z\hat s''_v\rangle + \langle \hat s’_x\hat s''_u\rangle-\langle \hat s’_x\hat s''_v\rangle$. 
The result,  $2\sqrt{2}$, violates Bell’s inequality  $\vert C\vert\le 2$ which would hold if the statistical properties of the four observables $\hat s’_z$, $\hat s’_x$, $\hat s''_u$, $\hat s''_v$ 
could be derived from a single ordinary probability distribution. There is {\it no joint probability distribution} governing {\it together} the four observables \cite{fine1982hidden}, 
although probabilities can be assigned separately to each pair $\hat s’_z\hat s''_u$, $\hat s’_z\hat s''_v$, $\hat s'_x\hat s''_u$, $\hat s’_x\hat s''_v$. 
 This  {\it irreparable} theoretical issue has been termed ``contextuality loophole'' \cite{nieuwenhuizen2011contextuality}.

There is also no joint probability distribution $P(x, p)$ for the two non-commuting observables $\hat x$ and $\hat p$  of a particle \cite{fine1982hidden}.
Heisenberg's inequality should not be regarded as a property of a single physical system, since the two $q$-variances $(\Delta x)^2$ and $(\Delta p)^2$ 
must be measured on different systems and hence cannot be interpreted as ordinary variances\footnote{The Wigner function $W(x, p)$, 
a mathematical representation of the density operator,  looks like a probability distribution without having its properties, since it may have 
negative parts. It is only by focusing on the single observable $\hat x$ (or $\hat p$) that one gets an ordinary probability distribution on $x$ (or $p$), represented by the integral of $W(x, p)$ over $p$ (or over $x$), and which produces independently $\Delta x$ (or $\Delta p$).  }.


The role of the apparatus is also emphasized by the consideration of {\it imperfect measurements}, which may be studied as exercises. (i)  If S and M are not sufficiently coupled (\cite{allahverdyan2013understanding},  §  7.3), truncation takes place but the source $\hat h_i$ entering the dynamical equation (\ref{eq4.5}) for $\hat R_{ii}(t)$ is too weak to ensure relaxation of M towards $\hat R_i$ and  {\it registration cannot take place}.
An admixture of another $\hat R_j$ ($j\neq i$) spoils the one-to-one correspondence between $\hatri$ and $\hat R_i$ in the final equilibrium state $\hat D(\tf)$ of S+M.
(ii) If the conservation laws $ [\hat H  , \hat \pi_i] = 0$ are not satisfied (Sec. 8.2 of  \cite{allahverdyan2013understanding}), the final state of S+M still has an equilibrium form 
\BEQ \label{eq6.7}
\hat D(\tf) = \sum_i q_i \hat \rho_i \otimes \hat R_i , 		
\EEQ
as in § \ref{sec4.3.1}, but $q_i$ differs from $p_i$ and $\hat \rho_i$ differs from $\hatri$. In both cases, the properties (\ref{eq6.4}) and (\ref{eq6.6}) of ideal measurements are violated.

\renewcommand{\thesubsection}{\thesection\arabic{subsection}}
\renewcommand{\thesubsubsection}{\thesubsection.\arabic{subsubsection}}
\subsection{Conceptual structure}
\renewcommand{\thesubsection}{\arabic{subsection}.}

\label{sec1.2}

Among the many existing approaches to measurement theory, we have presented above ideal measurements as dynamical processes. An abstract formalism describing 
a set $\cE$ of systems S+A has eventually produced a physical interpretation. Our successive steps comply with the structure of a physical theory, 
as epistemologically analyzed by Darrigol  \cite{darrigol2015some} who distinguishes four ingredients:

\begin{itemize}
\item{(a)} A {\it ``symbolic universe''} made of mathematical entities and structures, associated with the description of the objects under study.

\item (b) Basic {\it theoretical laws} which govern, in this symbolic universe, the behavior of the abstract mathematical images of physical objects.

\item (c) {\it Interpretive schemes} which relate the above formalism to idealized physical experiments. Some mathematical quantities thereby acquire an interpretation and become meaningful in concrete experiments.

\item (d)  {\it Approximations} used to adequately set up a symbolic universe and theoretical laws, then to implement this mathematical formalism  
into the analysis of interpretive idealized experiments.
 
 \end{itemize}
 
 These four points underlie the above developments. The abstract quantum formalism of § \ref{sec2.1}  and § \ref{sec2.2} encompasses the points (a) and (b), respectively. 
 We use the term ``interpretation'' in the sense of point (c): Rather than trying to interpret the quantum formalism itself, we define a model of ideal quantum measurements (§ \ref{sec3}) 
 then formally analyze their process (§ \ref{sec4}). 
 
 Approximations (point (d)) are needed, as usual in statistical physics, to solve the equations that govern the dynamics of the compound system S+A, and to explain irreversibility. Moreover, {\it the interpretation itself} of a measurement process relies on approximations (Appendix \ref{secC}). In particular, the apparent incompatibility between von Neumann's reduction and Schr\"odinger's dynamics is elucidated owing to approximations which produce, on physically relevant timescales, negligible errors for a large enough apparatus.

 In order to overcome the remaining measurement problem, we needed to introduce the postulate $\cP$ which allowed us to theoretically describe the subsets $\cE_i$ 
 of $\cE$ (§ \ref{sec5}). The property $\cP$ appears as a ``{\it module}'' as defined by Darrigol  \cite{darrigol2015some}. He notes that incorporating in a new theory a ``module'', that is, some element of a former theory, is a general procedure which may ensure the compatibility, flexibility and intelligibility of both theories. Here, the purely quantum principles of § \ref{sec2} are complemented by the ingredient $\cP$ borrowed from pre-quantum physics and supported by experiment, which accounts for the ``Heisenberg cut''.

 This addition to the formal principles of § \ref{sec2} has allowed us to set up an interpretive scheme as defined by the above point (c), in which the apparatus 
 plays a major role (§ \ref{sec6}).

\renewcommand{\thesection}{\arabic{section}}
\section{Tutorial hints}
\renewcommand{\thesection}{\arabic{section}.}

\label{sec7}

The above material is an account of detailed courses that have been delivered to {\it master and graduate students or young researchers}, in the framework of doctoral studies or summer schools. At this level, the main text should provide a scheme for more or less thorough series of lectures. They may range from a clear formulation of the measurement problem to its proposed solution. For more advanced audiences, possibly questionable points may be discussed, such as the need of approximations to justify the truncation of off-diagonal blocks, or the status of the postulate $\cP$ facing the standard quantum principles. Exercises are suggested throughout.

Parts of the text may, however, inspire other types of courses.

In {\it quantum computing} or {\it communication}, the data are processed in the form of hidden $q$-information, and extraction of ordinary information therefrom is akin to a measurement process. Thus, many ideas issued from measurement theory fit in courses on quantum information. For instance, it is interesting to grasp the physical process of truncation that causes the destruction of some $q$-information, a property needed to read out some usable information (§ \ref{sec6.3} and Appendix \ref{secB}). 

Students in {\it epistemology} having some notions of quantum physics may benefit from lessons, adapted to such an audience, 
about the conceptual structure of the theory and about its physical interpretation (§ \ref{sec6} and Appendix \ref{secC}).

Many topics related to measurement processes are suited to illustrate courses of {\it statistical physics} at undergraduate or graduate level, either as lectures or as exercises (§ \ref{sec4.2} and Appendices \ref{secA}, \ref{secB}, \ref{secC}). When introducing density operators (formally as in § \ref{sec2}), it is advisable to discuss Schr\"odinger's ambiguity to clarify their physical meaning (§ \ref{sec5} and Appendix \ref{secD})  \cite{park1968nature}. Assignment of a quantum state to a system in thermodynamic equilibrium, including broken invariance, appears in §§ \ref{sec2.3}, \ref{sec3.2.1}, \ref{sec4.3.1} and \ref{sec5.3}. The role of conservation laws is exemplified in §§ \ref{sec3.2.2}, \ref{sec4.1}, \ref{sec4.3.2} and \ref{sec5.3}. Appendix \ref{secB} presents a pedagogical example of cascade relaxation mechanism, which destroys information about the observables of S incompatible with the tested quantity $\hat s$. This model helps to elucidate the irreversibility paradox (Appendix \ref{secC}).

      At an elementary level, it is essential to avoid misunderstandings when introducing the quantum concepts and formalisms. Ideal measurements play a major role in the interpretation of quantum mechanics. Therefore, although their complete theory is not accessible to beginners, some ideas extracted from § \ref{sec2} and § \ref{sec6} should be instilled in {\it high school} teaching or in {\it undergraduate courses}. Students should be told at the earliest possible stage that the quantum formalism is {\it abstract}, and that a wave function is not a physical property of a particle or a system. 
      A ``quantum state'' (even pure) does not describe an individual system, but is a mathematical tool for making probabilistic predictions.
      In this prospect, we advocated the introduction of the terms $q$-probabilities, $q$-correlations, etc., instead of probabilities, correlations, etc. to distinguish such quantum formal quantities from ordinary ones (§ \ref{sec2.1.3}), a distinction which is essential to discuss specifically quantum properties arising from the non-commutation of observables. Physical meaning emerges from such a formalism only in relation with a specified {\it context} or 
      a dedicated {\it apparatus} (§ 6.4). This feature should provide a proper understanding of Born's rule which arises from the indications of the pointer (§ 6.2), and of von Neumann's  ``collapse'' (§ 6.3) which is not a dynamical property but an updating of the quantum state based on a selection by the experimenter of one of the possible outcomes. (This demystifying idea is also suitable for popularization of science.)

We close by noting that ideas of quantum measurement theory find applications in various interdisciplinary areas, beyond quantum physics  \cite{Khrennikov2010ubiquitous,Khrennikov2015quantum,korbicz2021roads}.

\section*{Appendices}

\setcounter{section}{0}
\renewcommand{\thesection}{\Alph{section}}

\section{Effects of the bath}
 \renewcommand{\thesection}{\arabic{section}}
 \renewcommand{\thesubsection}{\Alph{section}\thesubsection}
\label{AppendixA}
\label{secA}

To fully establish measurement theory on the principles of quantum mechanics, one should solve the dynamical equations (\ref{eq4.5}) for M + B,  {in the respective eigensectors of S. 
A standard step in statistical mechanics consists in eliminating the bath B from (\ref{eq4.5}),  at second order in the weak coupling $\hat H_\rmMB$, so as to derive from (\ref{eq4.5}) approximate equations for M (see Sec. 4 of  \cite{allahverdyan2013understanding}). Besides the terms exhibited in (\ref{eqB.1}) below, these dynamical equations for $\hat R_{ij}(t)$ include a retarded non-Hamiltonian part issued from $\hat H_\rmMB$. 
They govern {\it solely the measuring device} M owing to the conservation of $\hat s$ which allows the splitting of the equations 
for $\hat D(t)$ (as in § \ref{sec4.1} for $\hat\cD$).

.
			
For $i = j$ (``{\it registration}''), the equation for $\hat R_{ii}(t)$ is exactly the same as for an apparatus M {\it alone} with Hamiltonian $\hat H_\rmM  +\hat h_i$, weakly coupled to a bath at temperature $T$. The relaxation of $\hat R_{ii} (t)$, from $\hat R(0)$ to the canonical equilibrium state $\hatRih $, taken for granted in § \ref{sec3.2.2}  (Eq. (\ref{eq3.7})) and § \ref{sec4.3}, can then be justified by applying standard techniques of statistical mechanics to measurement models. For the Curie-Weiss model, a detailed study, including a discussion of imperfect measurements, is presented in Secs. 7 and 8. of  \cite{allahverdyan2013understanding}.

For $i\neq j$, the decay of $\hat R_{ij}(t)$ (``{\it truncation}'') may result from a {\it decoherence} induced solely by the bath.
This occurs for many models (reviewed in  \cite{allahverdyan2013understanding}, Sec. 2) which can inspire exercises. 
The bath terms may also suppress the recurrences exhibited below in Eq. (\ref{eqB.7}) (\cite{allahverdyan2013understanding},  §   6.2). 
However, as illustrated by appendix B, the decay of $\hat R_{ij}(t)$ may also take place without a bath.

\renewcommand{\thesection}{\Alph{section}}
 \setcounter{section}{1}
 \setcounter{subsection}{0}
 \section{Truncation}
\renewcommand{\thesection}{\Alph{section}}
 \renewcommand{\thesubsection}{\Alph{section}.\arabic{subsection}}
\label{AppendixB}
\label{secB}

Among the various mechanisms that can ensure the truncation of $\hat \cD(t)$ (§ \ref{sec4.2}), we focus here on {\it destructive interferences due to dephasing}, which are efficient even {\it in the absence of a bath}. The dynamical equations for M then reduce to 
\BEQ \label{eqB.1}
\rmi\hbar  \frac{\d\hat R_{ij}(t)}{\d t} = [\hat H_\rmM , \hat R_{ij}(t)] + \hat h_i \hat R_{ij}(t) - \hat R_{ij}(t) \hat h_j ,     
\EEQ
with initial condition $\hat R_{ij}(0) = \hat R_\rmM (0)$. The decay of  $\hat R_{ij}(t)$ will rely on the large number of degrees of freedom of the pointer, and on the non-Hamiltonian form of (\ref{eqB.1}). For the Curie-Weiss model, this mechanism has been thoroughly explored in Sections  \ref{sec5}  and \ref{sec6} of  \cite{allahverdyan2013understanding}, including discussions about the needed size of parameters, the time scales, the time dependence of correlations, the validity of approximations, the role of the initial state of M, the possibility of recurrences and their suppression, the interplay with the decoherence by the bath. Such developments lend themselves to students' exercises which enlighten the irreversibility of truncation. We suggest simple ones below.

\renewcommand{\thesubsection}{\thesection.\arabic{subsection}}
\renewcommand{\thesubsubsection}{\thesubsection.\arabic{subsubsection}}
\subsection{Dephasing mechanism}
\renewcommand{\thesubsection}{\arabic{subsection}.}
\label{appB.1}
\label{secB.1}

If all operators $\hat H_\rmM $, $\hat h_i$ and $\hat R_\rmM (0)$ commute with one another, Eq. (\ref{eqB.1}) is solved as 
\BEQ \label{eqB.2}
\hspace{-7mm}	
\hat R_{ij}(t) = \hat R_\rmM (0) \exp [- \rmi (\hat h_i - \hat h_j) {t }/{\hbar} ] . 
\EEQ
For $i\neq j$, the existence of many degrees of freedom for M allows $\hat h_i - \hat h_j$ to have a large number of different eigenvalues. Their contributions to ${\rm tr}_\rmM  \hat R_{ij}(t) \hat O$ for ``smooth'' observables $\hat O$ may then oscillate out of phase, and thus {\it interfere destructively}, so that such $q$-expectation values will  decay rapidly (\cite{allahverdyan2013understanding}, subsection 5.1.2).

The Curie-Weiss model provides a simple illustration. The registering device M is modelled by an Ising magnet involving $N$ spins $\half$  with $z$-components $\hat \sigma_z^{(n)}= \pm 1$ ($1\leq  n \leq  N$).
 \footnote{As Pauli matrices, the spin operators have the representation
 $\hat\sigma_x^{(n)}=\left(\!\!\begin{array}{ll} 0 & 1 \\  1 & 0 \end{array}\!\!\right)$, 
 $\,\hat\sigma_y^{(n)}=\left(\!\!\begin{array}{ll} 0 & \!\!  \! -i \\  i & 0 \end{array}\!\!\right)$, 
 $\,\hat\sigma_z^{(n)}=\left(\!\!\begin{array}{ll} 1 & 0 \\  0 &\! \! \! -1 \end{array}\!\!\right)$, and similarly for $\hat s_x$, $\hat s_y$, $\hat s_z$.\label{Paulimatrices}}
 The pointer observable $\hat A$  {in the $2^N$-dimensional Hilbert space ${\cal H}_{\rm M}$}  is the magnetization operator
$\hat M_z= \sum _{n=1}^N \hat \sigma_z^{(n)}$, which takes macroscopic values $+M_\rmF$ or $-M_\rmF$ in the ferromagnetic equilibrium states $\hat R_\Uparrow$  
or $\hat R_\Downarrow$,  with fluctuations of order $\sqrt{N}$.  At $t=0$, M lies in the metastable paramagnetic state $\hat R_\rmM (0) =2^{-N} \hat I$, 
with which the Ising Hamiltonian $\hat H_\rmM $ commutes. We consider a situation in which the truncation time is much shorter that the registration time, 
so that the diagonal blocks $\hat R_{\up\up}(t)$ and $\hat R_{\down\down}(t)$ remain practically equal to $\hat R_\rmM(0)$ during this truncation process.  

 The system S is a spin $\half$  with Pauli matrices $\hat s_x$, $\hat s_y$, $\hat s_z$, the tested observable $\hat s$ being $\hat s_z=\hat \pi _\uparrow -\hat \pi_\downarrow$. 
 (Ref. \cite{nieuwenhuizen2022models} considers the situation for spins greater than $\half$.)
 The index $i$ takes two values, denoted by $\uparrow$  and $\downarrow$ or $\pm1$, and the initial state of S is represented by a $2 \times 2$ density matrix $\hat r(0$). We take as interaction Hamiltonian 
\BEQ  \label{eqB.3}
    \hat H_\rmSM &=& \hat \pi _\uparrow  \otimes \hat h  _\Uparrow +\hat \pi_\downarrow  \otimes \hat h_\Downarrow
= -\sum_{n=1}^N g \hat s_z \otimes\hat \sigma_z^{(n)} = -g\hat s_z \otimes\hat M_z 
\EEQ
with $g>0$, which favorizes a pointer value having the same sign as $s_z$. The sources $\hat h_i$ in (\ref{eqB.1}) are 
\BEQ  \label{eqB.4}
\hat h_\Uparrow  = -\hat h_\Downarrow  = -g\sum_{n=1}^N  \hat \sigma_z^{(n)},	\EEQ
which act as external fields $\pm g$. The solution (\ref{eqB.2}) of (\ref{eqB.1}) is 
\BEQ  \label{eqB.5}
\hat R_{\uparrow\downarrow }(t) = \prod_{ n=1}^N \left(\half  \exp \frac{2\rmi g \hat \sigma_z^{(n)} t }{\hbar}\right) = \hat R_{\downarrow \uparrow}^\dagger  (t) ,     
\EEQ
and it yields for the state of S+M (note that $(\pi _\uparrow)_{ij}=\delta_{i,1}\delta_{j,1}$ and  $(\pi _\downarrow)_{ij}=\delta_{i,-1}\delta_{j,-1}$,
while  $\hat s_\pm =\half (\hat s_x \mp  i\hat s_y)$):
\BEQ \label{eqB.6} \hspace{-5mm}
\hat D(t) &=& [r_{\uparrow\uparrow} (0)\hat \pi _\uparrow  + r_{\downarrow\downarrow} (0)\hat \pi_\downarrow ] \otimes 
\frac{1}{2^N} \hat I +       r_{\uparrow\downarrow} (0) \hat s_+ \otimes\hat R_{\uparrow\downarrow} (t) 
+ r_{\downarrow \uparrow } (0)\hat s_- \otimes\hat R_{\downarrow\uparrow}  (t).       
\EEQ

The oscillations exhibited in (\ref{eqB.5}) involve $2^N$ frequencies, the spectrum of which extends  uniformly from $-Ng/\pi \hbar$  to $Ng/\pi \hbar$. In the calculation of the $q$-expectation values $\langle \hat s_x(t) \rangle$  and $\langle \hat s_y(t) \rangle$  associated with the marginal state $\hat r(t) = {\rm tr}_\rmM \hat D(t)$ of S, these oscillations interfere and yield
\BEQ \label{eqB.7}
	\langle \hat s_x(t) \rangle  = {\rm Tr}_\rmSM \hat D(t) \hat s_x = \langle \hat s_x(0) \rangle  \cos^N \frac{2gt}{\hbar} 
\EEQ
(and likewise with $x\mapsto y$ for $\langle \hat s_y(t) \rangle$). At the beginning of the process, for $t \ll  \pi \hbar /4g$, we have 
$\cos 2gt/\hbar  \approx \exp(-2g^2t^2/\hbar ^2)$. Hence,
\BEQ \label{eqB.8}
\langle \hat s_x(t) \rangle  \sim \langle \hat s_x(0) \rangle  e^{-t^2/\tau ^2} , \quad  \tau ^2 \equiv  \frac{\hbar ^2}{2Ng^2  },    
\EEQ
fades out on the time scale $\tau$, which is {\it very short} for large $N$. Thus, at the very first stage of the interaction of S with the apparatus, the measurement of $\hat s_z$ destroys through $\hat H_\rmSM$ the original information about the {\it observables of} S  {\it that do not commute with} $\hat s_z$, which are originally embedded in $r_{\downarrow \uparrow}  (0) = \half \langle \hat s_x(0) \rangle +\frac{i}{2} \langle \hat s_y(0) \rangle$ and $r_{\uparrow \downarrow}  (0) =[r_{\downarrow \uparrow}  (0)] ^\ast$.

\renewcommand{\thesubsection}{\thesection.\arabic{subsection}}
\renewcommand{\thesubsubsection}{\thesubsection.\arabic{subsubsection}}
\subsection{Possible recurrences}
\label{appB.2}
\label{secB.2}

 The above model suffers from a defect. Although truncation is achieved soon after the beginning of the process, the factor $\cos^N2gt/\hbar$  in (\ref{eqB.7}) is periodic, and gives rise to {\it recurrences}. Long after they have disappeared, $\langle \hat s_x(t) \rangle$  and $\langle \hat s_y(t)\rangle$  return to their initial values (within a sign for odd $N$) at the times $t_\nu  = \nu \pi \hbar /2g \gg \tau$  ($\nu  =1, 2, 3,\cdots$), around which they display narrow gaussian peaks. These recurrences originate from a peculiarity of (\ref{eqB.4}): the eigenvalues of $\hat h_\Uparrow  - \hat h_\Downarrow$  have the same spacing $4g$ and a large degeneracy.  A more uniform spectrum can let such recurrences disappear. 
 
To illustrate this fact, we improve the model by assuming that the couplings of the tested spin $\hat s$ with the spins $\hat \sigma_z^{(n)}$ of the pointer are not exactly the same (subsection 6.1 of  \cite{allahverdyan2013understanding}). This dependence substitutes $g_n = g + \delta g_n$ to $g$ in (\ref{eqB.3}), (\ref{eqB.4}) and (\ref{eqB.5}). The deviations $\delta g_n = g_n - g$ are assumed to be small ($\delta g_n\ll g$) and to have a zero average. 
Accordingly, this spread has a negligible effect on the dynamics of the diagonal blocks $\hat R_{ii}(t)$,  i.e., on the registration process.

In (\ref{eqB.7}), $\cos^N 2gt/\hbar$  is replaced by $\prod_{ n=1}^N \cos 2g_nt/\hbar$. The factors of this product are now out of phase, and for $t = t_\nu  = \nu \pi \hbar /2g$, each one reduces to $(-1)^\nu \cos(\nu \pi \delta g_n/g)$. The shape of the peak around $t_\nu$  remains nearly unchanged, but its height is strongly reduced. Denoting by $\delta g^2$ the mean square of the $\delta g_n$'s, and assuming that $\nu \delta g/g \ll 1$, we find for $t = t_\nu$  the factor 
\BEQ  \label{eqB.9}
 && \hspace{-10mm}
\prod_{n=1}^N \cos \frac{ \nu\pi\delta g_n}{g} = \exp \sum_{n=1}^N\ln  \cos\frac{ \nu\pi\delta g_n}{g} 
\sim \exp\left( -\sum_{n=1}^N \frac{ \nu^2\pi^2\delta g_n^2}{2g^2}  \right) = \exp \left(- N\frac{ \nu^2\pi^2\delta g^2}{2g^2} \right).
\EEQ
Hence, if the dispersion of the couplings $g_n$ satisfies $1\gg  \delta g/g \gg 1/\sqrt{N}$, the initial relaxation as $\exp(-t^2/2\tau ^2)$ is not modified, but the $\nu$'th peak is damped by a factor $\exp-K\nu ^2$, where $K\equiv N(\pi  \delta g/g)^2/2 \gg 1$. 
Thus, for a coupling $\hat H_\rmSM = -\sum_{n=1}^N g_n \hat s_z \otimes \hat \sigma_z^{(n)}$, 
the relaxation exhibited in (\ref{eqB.8}) is not spoiled by recurrences, and the {\it truncation} after the time $\tau$  
of the off-diagonal blocks of $\hat D(t)$ is {\it permanent}. \footnote{We focus here on not too large values of $\nu$, on the time range $t\ll \hbar /\delta g$. For larger values of $\nu$, i.e., of time, the irreversible truncation may be ensured by the bath. 
For this decoherence process after the dephasing, we find in subsection
 6.2 of  \cite{allahverdyan2013understanding}  a relaxation time $\tau\sim N^{-1/4}$ for large $N$.}

\renewcommand{\thesubsection}{\thesection.\arabic{subsection}}
\renewcommand{\thesubsubsection}{\thesubsection.\arabic{subsubsection}}
\subsection{Cascade of correlations}
\label{appB.3}
\label{secB.3}

 Returning to the state (\ref{eqB.5}), (\ref{eqB.6}) of S+M, let us now consider the $q$-expectation values of more and more complex observables $\hat s_x \otimes \hat \sigma_z^{(n)}$, $\hat s_y \otimes \hat \sigma_z^{(n)}$, $\hat s_x \otimes \hat \sigma_z^{(n)} \hat \sigma_z^{(n' )},\cdots$, astride S and M. As in the evaluation of (\ref{eqB.7}), we find for $t \ll   \pi \hbar /4g$,
\BEQ \label{eqB.10}
\langle \hat s_x \otimes \hat \sigma_z^{(n)}(t)\rangle  = \langle \hat s_x \otimes \hat m(t)\rangle  \sim 
\langle \hat s_y (0) \rangle  \sim \left(\frac{\sqrt{2}}{\sqrt{N}}\frac{t}{\tau} \right) e^{-t^2/\tau ^2}.   
\EEQ
Thus, the $q$-correlation\footnote{There is no subtractive term $\langle \hat s_x(t) \rangle \langle \hat \sigma_z^{(n)}(t)\rangle$  
since $\langle \hat \sigma_z^{(n)}(t)\rangle \approx 0$ on the truncation time scale.}
of $\hat s_x$ (or $\hat s_y$) with the reduced pointer variable $\hat m=\hat M/N$, which is absent in the initial state, 
{\it increases then disappears} on the short time scale $\tau$. This property should be contrasted with the behavior of 
$\langle \hat s_z\otimes\hat m(t)\rangle$  which characterizes ``registration'' (§ \ref{sec4.2}), a quantity which rises from $0$ to $M_\rmF/N$ 
between the times 0 and $\tf$ under the effect of the interaction of M with the bath B. 
               
The $q$-correlations between $\hat s_z$ and an increasing number $k$ of elements $\hat \sigma_z^{(n)}$ of the pointer (for different values of $n$, $n'$, $n''$) 
are likewise found, for $ t\ll  \pi \hbar /4g$, as
\BEQ \label{eqB.11}
\langle \hat s_x \otimes \hat \sigma_z^{(n)} \hat \sigma_z^{(n')}(t)\rangle  \sim - \langle \hat s_x(0) \rangle
  \left(\frac{\sqrt{2}}{\sqrt{N}}\frac{t}{\tau} \right)^2 e^{-t^2/\tau ^2},       	   
\EEQ
and
 \BEQ \label{eqB.12}
\hspace{-5mm}
\langle \hat s_x \otimes \hat \sigma_z^{(n)} \hat \sigma_z^{(n' )} \hat \sigma_z^{(n'')} (t)\rangle  \sim -
 \langle \hat s_y(0)   \left(\frac{\sqrt{2}}{\sqrt{N}}\frac{t}{\tau} \right)^3 e^{-t^2/\tau ^2} ,
\EEQ
and so on. While being absent at the time $t = 0$  (no initial correlation between the apparatus A and the tested system S), they begin to be produced after the interaction between S and M has been switched on. 
Initially, they grow as  $(t/\tau)^k$ with a small amplitude $\sim N^{-k/2}$, and, for increasing $k$, reach a lower and lower peak later and later, 
for $t/\tau = \sqrt{k/2}$, and then rapidly decay. 

	Altogether, the decay (\ref{eqB.8}) of $\langle \hat s_x(t) \rangle$  and $\langle \hat s_y(t) \rangle$  is accompanied by an appearance of $q$-{\it correlations} (\ref{eqB.10}) of $\hat s_x$ or $\hat s_y$ with each spin $\hat \sigma_z^{(n)}$ of M.  Then, while  the latter disappear in turn, they give way to $q$-correlations (\ref{eqB.11}) of $\hat s_x$ or $\hat s_y$ with two spins of M, then with three (\ref{eqB.12}), and so on. Larger and larger numbers $k$ of degrees of freedom of the pointer are gradually involved, at later and later times $t \sim \tau  \sqrt{k/2}$, in a {\it cascade phenomenon}; their $q$-correlations with S successively build up a little before fading out
 \footnote{The $q$-correlations with large values of $k$ which appear over times $t\gg  \tau$  are discussed in subsections 5.1.3 and 5.3.2  of ref.  \cite{allahverdyan2013understanding} \label{fnB.3}}
		
Thus, the $q$-information about the observables incompatible with the measured quantity $\hat s_z$, originally embedded in $\hat r(0$) as $\langle \hat s_x(0) \rangle$  and $\langle \hat s_y(0) \rangle$, {\it flows step by step} towards more and more complicated $q$-correlations between S and M, which eventually become physically unattainable and can be disregarded.  Altogether, the {\it irreversibility of von Neumann's reduction} is understood as a discarding of the inaccessible
multiparticle $q$-correlations between the ``cat'' operators $\hat s_{x}$, $\hat s_y$ of S and the spins of the macroscopic pointer, 
which are generated by the dynamics as the result of a cascade.

\renewcommand{\thesection}{\Alph{section}}
\section{Conceptual need for approximations}
\setcounter{figure}{0}
\renewcommand{\thesection}{\Alph{section}}
\label{AppendixC}
\label{secC}

A quantum measurement, which involves the interaction of the tested system with a macroscopic apparatus, is an irreversible process, 
and its analysis raises the {\it irreversibility paradox}. This irreversibility of the process  seems to contradict the unitarity of the evolution
 generated by the Liouville–von Neumann equation.
An especially simple illustration is provided by the model of App. \ref{secB}. The exact solution (\ref{eqB.2}) of the equation of motion 
yields a constant entropy $S[\hat D(t)]$, and also 
satisfies $\hat R_{\uparrow\downarrow} (t) \hat R_{\uparrow\downarrow}^\dagger  (t)= 2^{-2N}\,\hat I$ at all times, {\it without any decay}
whereas the off-diagonal blocks $\hat R_{\uparrow\downarrow} (t)$ must {\it effectively disappear} physically after some time, 
a property needed to explain the reduction.  Such mathematical properties cannot be tested experimentally, hence they have 
no physical meaning; in fact, their presence stresses once more the need for approximations (item (d) in §   \ref{sec1.2}).
 
  The treatment of App. \ref{secB} elucidates this apparent contradiction. Under reasonable conditions, all physical quantities (\ref{eqB.8}), 
   (\ref{eqB.10}) --  (\ref{eqB.12}) vanish for $t\gg  \tau$  as if the off-diagonal blocks $\hat R_{\uparrow\downarrow} (t)$ and $\hat R_{\downarrow \uparrow}  (t)$ were chopped off at such times (in spite of the property $\hat R_{\uparrow\downarrow} (t) \hat R_{\uparrow\downarrow}^\dagger  (t)= 1/2^{2N}$).  As usual when reconciling statistical mechanics with thermodynamics, this solution of the irreversibility paradox relies on two key points. 
   
On the one hand, the ``state'' (\ref{eqB.5}), (\ref{eqB.6}) is a {\it mathematical} object which encompasses the probabilistic predictions that we can make about any physical quantity (end of § \ref{sec2.1.4}). However, when $t$ increases, this exact state contains more and more {\it physically irrelevant information} about observables that correlate a very large number $k$ of constituents 
(App. B.3). \ref{appB.3}). 
As such quantities cannot be grasped, neither experimentally nor theoretically, it is legitimate to discard them.

On the other hand, the decays of (\ref{eqB.8}), (\ref{eqB.10}) -- (\ref{eqB.12}), that lead to truncation, originate from the large number $N$ of degrees of freedom of the {\it macroscopic system} M, resulting in the huge number $2^{2N}$ of  complex  elements of the $2^N\times 2^N$ matrix $\hat R_{\uparrow\downarrow} (t)$. 
These approximate decays become mathematically exact in the limit $N\to \infty$ and fixed $g\sqrt{N}$, for finite values of the time $t$ and of the number $k$ of spins involved in the $q$-correlations. 

Another essential point lies in the consideration of {\it time scales}. 
In App. \ref{appB.1}, 
the decay takes place on the time scale $\tau  = \hbar /\sqrt{2N}g$, and it is effective for $t\ll  \pi \hbar /4g$. In the less simplified model of App. 
\ref{appB.2}, 
recurrences are eliminated on a much larger time scale $\hbar /\sqrt{2N}\delta g$ (with $\delta g/g \gg 1/\sqrt{N}$). The comparison between App. 
\ref{secB.1}  and \ref{secB.2} also shows that sufficiently {\it complicated interactions} are needed to ensure irreversibility of the cascade process 
and unreachably {\it large ``Poincar\'e''  recurrence times}.

All the above features are general. The quantum dynamical equations (\ref{eq2.2}) or (\ref{eq4.5}) refer to {\it finite} systems. They are reversible and their {\it exact} solution cannot provide the result (\ref{eq4.16}) for $\hat \cD(\tf)$. However, we have stressed (§ \ref{sec3.1} and § \ref{sec4.3}) that both the measuring device M and the bath B must be {\it large}. 
Then, suitably chosen approximations provide negligible errors for the physically relevant quantities in the limit of a large apparatus, and the use of (\ref{eq4.16}) is justified. As in App. 
\ref{secB.3}, interactions gradually transfer, within $\hat\cD(t)$, physically accessible $q$-information towards elusive $q$-correlations involving a larger and larger number of degrees of freedom, which are discarded to yield the physical result (\ref{eq4.16}) for $\hat \cD(\tf)$.

 This loss of irrelevant $q$-information is measured by the increase $S[\hat \cD(\tf)] - S[\hat \cD(0)]$   \cite{balian2005information}.
The constancy of the von Neumann entropy $S[\hat \cD(t)]$, which results from the exact dynamical equation (\ref{eq2.2}), appears as a mathematical property without physical incidence,
 as in the irreversibility paradox, since the disregarded variables are unreachable. 
 
Besides making calculations feasible, approximations are required for conceptual reasons (§ \ref{sec1.2}, point (d)) \cite{darrigol2015some}. 
In fact, the {\it exact state} $\hat \cD_{\rm exact}(\tf)$ issued from $\hat \cD(0)$ through the equation of motion (\ref{eq2.2}) {\it cannot be given any interpretation}. 
 Strictly speaking, $\hat \cD_{\rm exact}(t)$ is a sum of oscillatory terms and has no limit. 
Two types of {\it interpretive approximations} enter the derivation of the physical $\hat \cD(\tf)$, which relies on independent relaxations of the blocks $\hat \cR_{ij}(t) = \hat \pi_i\hat \cD (t)\hat \pi_j$ of $\hat \cD(t)$. For $j = i$, M evolves irreversibly towards an equilibrium state $\hat R_i={\rm tr}_\rmB \hat \cR_i$ characterized by the outcome $A_i$ of the pointer, a property needed to ensure the {\it registration}, i.e., the one-to-one relationship between $A_i$ and $s_i$ (§ \ref{sec4.2}). For $j\neq  i$, while the exact equations (\ref{eq4.5}) provide ${\rm Tr}_A \hat \cR_{ij}(t) \hat \cR^\dagger_{ij}(t) = [{\rm Tr}_\rmM  \hat R_\rmM ^2(0)][ {\rm Tr}_\rmB  \hat R_\rmB ^2(0)]$, an approximation is needed to explain the {\it truncation}, i.e., the 
 {\it effective}  decay of $\hat \cR_{ij}(t)$ towards $\hat \cR_{ij}(\tf) = 0$. Without this property, we could not describe in the quantum framework the splitting of the set $\cE$  
into subsets $\cE_i$, since we could not split $\hat \cD(\tf)$ into a sum of terms $p_i \hatri \otimes\hat R_i \otimes\hat R_\rmB (0)$. 

Thus, approximations, justified in the limit of a large apparatus, appear as essential ingredients in the very {\it elaboration of an interpretation} of ideal measurement processes.

\renewcommand{\thesection}{\Alph{section}}
 \setcounter{section}{3}
 \setcounter{subsection}{0}
 \section{Ambiguity of decompositions of density operators}
\renewcommand{\thesection}{\Alph{section}}
 \renewcommand{\thesubsection}{\Alph{section}\arabic{subsection}}
\label{AppendixD}
\label{secD}

	We will prove that, if a non-pure density operator $\hat \cD$ is expressed as a weighted sum $\sum_k \lambda_k \hat \cD_k$, the {\it separate} terms of this decomposition {\it cannot be given a physical meaning} if nothing else than $\hat \cD$ is known. As in the example discussed in § \ref{sec5.1}, it will suffice to exhibit two such decompositions that give rise to a logical contradiction  \cite{park1968nature}.

\renewcommand{\thesubsection}{\thesection.\arabic{subsection}}
\renewcommand{\thesubsubsection}{\thesubsection.\arabic{subsubsection}}
\subsection{Two-dimensional density operator}	
 
We begin with an arbitrary state $\hat \cD$ in a 2-dimensional Hilbert space $\cH^{(2)}$, which formally describes a set $\cE$ of spins $\half$  or of $q$-bits. Expressing the density matrix $\hat \cD$ on the basis of Pauli matrices as$^{\ref{Paulimatrices}}$  
\BEQ \label{eqD.1}
\hat \cD \equiv  \half  (\hat I + v_x \hat \sigma_x +v_y \hat \sigma_y + v_z \hat \sigma_z), 	
\EEQ
we visualize it as the end point of the 3-dimensional vector ${\bf v}$ which represents the polarization. 
(To simplify, we shall term ``vector''  ${\bf v}$ the end point of the vector ${\bf v}$.) 
For a pure state $\hat \cD=\,| \psi \rangle \langle \psi |$, the vector  ${\bf v}$ lies on the Poincar\'e--Bloch sphere $| {\bf v}| =1$, and we have 
$| \langle \psi_1| \psi_2\rangle | = \half  (1+ {\bf v}_1\cdot  {\bf v}_2$);  for a mixed state,  in which case $| {\bf v}| < 1$, ${\bf v}$ lies inside the sphere. 
  
 If we consider two (or more) disjoint sets $\cE_k$ of $\cN_k$ spins, represented by states $\hat\cD_k$, their merger produces a set $\cE$ of $\cN =\sum_k \cN_k$ spins, 
 represented, according to (\ref{eq2.3}), by the state $\hat\cD$ given by $\cN \hat\cD =\sum_k \cN_k \hat\cD_k$. In the Poincar\'e--Bloch formalism, 
 the point ${\bf v}$ associated with $\hat\cD$ is the {\it barycenter} (defined by $\cN{\bf v} =\sum_k \cN_k {\bf v}_k$) of the points ${\bf v}_k$ associated with $\hat\cD_k$.
Conversely, consider a mixed state $\hat\cD$ represented by the point ${\bf v}$ within the unit sphere. It has a mathematical decomposition 
\BEQ  \label{eqD.2}
\hat \cD = | \psi _1\rangle  \rho_1 \langle \psi_1| +| \psi_2\rangle  \rho_2 \langle \psi_2|
\quad    \textrm{      or     }  \quad    {\bf v} = \rho_1{\bf v}_1 + \rho_2{\bf v}_2	 ,
\EEQ
 (with positive real numbers $ 0 < \rho_{1,2} <  1$,  $\rho_1+ \rho_2=1$), visualized by a straight line which passes through ${\bf v}$ and crosses the unit sphere at two points ${\bf v}_1$ and ${\bf v}_2$. Let us consider another decomposition ${\bf v} = \rho '_1{\bf v}'_1 + \rho '_2{\bf v}'_2$. The two points ${\bf v}_1$ and ${\bf v}'_1$, and hence the corresponding kets $| \psi _1\rangle$ and $| \psi '_1\rangle$, are different. If $| \psi_1\rangle$  did describe a physical subset $\cE_1$ of $\cE$ and $| \psi '_1\rangle$  another subset $\cE'_1$, there would exist a non-empty physical subset $\cE_1 \cap \cE'_1$ simultaneously described by two different states $| \psi_1\rangle$  and $| \psi '_1\rangle$. Such a contradiction arises for any pair of decompositions. Hence, decompositions of $\hat \cD$ 
cannot have a physical meaning. (As indicated in § \ref{sec5.3}, additional information, not included in $\hat \cD$, 
is needed to interpret one among the various decompositions.) 
 
 The two pure states entering (\ref{eqD.2}) are orthogonal ($\vv_1 = -\vv_2$) for the unpolarized state $\vv = 0$. Otherwise, for $|\vv| \neq 0$, imposing  $\langle\psi_1|\psi_2\rangle=0$ in (\ref{eqD.2}) would provide only one decomposition of $\hat\cD$. However, there is no reason to assume such an orthogonality, since $\hat\cD$ may have been prepared by a merger $\cE$ of arbitrary sets $\cE_k$ with a loss of information about this construction.  Moreover, this assumption would produce an unphysical qualitative difference between states described by $\vv = 0$ (infinity of decompositions) and by the limit $|\vv|\to 0$ (single decomposition).

We have split above $\hat \cD$ into two pure states, which cannot be split further. Other decompositions of $\hat \cD$ may be considered, for which $\vv$ is the barycenter of an arbitrary number of vectors  $\vv_k$ lying either on the surface of the unit sphere or inside it. Their ambiguity follows from that of the elementary decompositions of the type (\ref{eqD.2})  \cite{park1968nature}. In case the set $\cE$ described by $\hat \cD$ had been built by randomly mixing pure or non-pure states prepared beforehand, it is impossible to retrieve from $\hat \cD$ these original states.

\renewcommand{\thesubsection}{\thesection.\arabic{subsection}}
\renewcommand{\thesubsubsection}{\thesubsection.\arabic{subsubsection}}
\subsection{Arbitrary dimensions}
\label{secD.2}
   
In the general case of a $n$-dimensional Hilbert space $\cH^{(n)}$, an arbitrary density operator $\hat \cD$ can be diagonalized as
\BEQ  \label{eqD.3}
\hat \cD = \sum_{k=1}^n | \phi_k\rangle  q_k \langle \phi_k|  		
\EEQ
over an orthonormal basis $| \phi_k\rangle$  of $\cH^{(n)}$. If $\hat\cD$ is not pure, at least two of its eigenvalues, say $q_1$ and $q_2$, do not vanish, and $0 <  q_1 + q_2 \equiv  q \leq  1$. Isolating the first two terms of (\ref{eqD.3}), we rewrite $\hat \cD$ as
\BEQ  \label{eqD.4}
	\hat \cD = q \hat \cD_2 + \sum_{ k=3}^n | \phi_k\rangle  q_k \langle \phi_k|  ,   	
\EEQ
where $\hat \cD_2 \equiv \,  | \phi_1\rangle  (q_1/q) \langle \phi_1|  + | \phi_2\rangle ( q_2 /q) \langle \phi_2 | $ has the same properties as a density operator in the 2-dimensional Hilbert subspace $\cH^{(2)}$ of $\cH^{(n)}$. The ambiguity of $\hat \cD_2$ entails that of $\hat \cD$.

The quantum ambiguity of the decompositions of density operators contrasts with the properties of classical probability distributions  \cite{park1968nature,allahverdyan2013understanding}. For a classical spin that may take the values $+1$ or $-1$ (or for an ordinary bit), the probability distribution $p_{+1}$, $p_{-1}$ can be visualized by a point  $x = p_{+1} - p_{-1}$ in the interval $[-1, +1]$ of a one-dimensional line, which replaces the 3-dimensional 
Poincar\'e -- Bloch sphere $|\vv|\le1$ of (\ref{eqD.1}). Then, the barycentric decomposition $x = p_{+1} - p_{-1}$ replacing (\ref{eqD.2}) is {\it the only one}, and there is no ambiguity. 
   
More generally, the density matrices in a  $n \times n$ Hilbert space can be geometrically visualized as the points $\vv$ of a ($n^2-1$)-dimensional convex region. The extreme points $x$ of this region, which lie in a ($2n-2$)-dimensional space, represent the pure states. A point $\vv$ associated with a non-pure state has an infinite, continuous number of barycentric representations in terms of extreme points $x$. In contrast, the ordinary probability distribution $p_1, p_2,\cdots p_n$ for $n$ exclusive events is visualized as a point $\vv$ in a convex ($n-1$)-dimensional region. This region (mathematically called ``regular simplex'') has 
$n$ discrete extreme points $x$ representing events known with certainty, in terms of which the barycentric representation of $\vv$ is unique.

\renewcommand{\thesubsection}{\thesection.\arabic{subsection}}
\subsection{Pure states versus mixed states}
    \label{secD.3}
         
         The mathematical distinction between pure and mixed states entails some physical differences. Given the state describing a set of systems, the above ambiguity of decompositions, which impedes to describe subsets, is specific to mixed states  \cite{park1968nature,allahverdyan2013understanding}.
         Regarding von Neumann's entropy \textcolor{red}{  $S$ }  as a measure of randomness (§ \ref{sec2.2.4}), pure states for which $S=0$ appear as the most informative, 
         the least random, or least ``noisy'' ones.

         We present below another physical criterion exhibiting {\it pure states} as the {\it least noisy} ones. Consider, in a $n$-dimensional Hilbert space $\cH$, a state $\hat r$ represented by a density matrix having $k$ non-zero eigenvalues ($1\le k\le n$). The projector $\hat\pi_r$ on the corresponding eigenvectors spans a $k$-dimensional Hilbert subspace that we denote by $\cH_r$. We shall call {\it ``dispersionless''} the observables $\hat a$ that have a vanishing $q$-variance on the state $\hat r$:
\BEQ  
\label{eqD.5}               \langle \hat a^2\rangle - \langle \hat a\rangle ^2 \equiv \Tr \hat r \hat a^2 - (\Tr \hat r \hat a)^2 = 0.               
\EEQ
We introduce the spectral decomposition $\hat a=\sum_i a_i \hat\pi_i$ of $\hat a$ in terms of its eigenvalues $a_i$ and eigenprojectors $\hat\pi_i$, and denote by $\cH_i$ the $m_i$-dimensional Hilbert subspace of $\cH$ spanned by $\hat\pi_i$. The condition (\ref{eqD.5}) reads
        \BEQ   \sum_i  p_i (a_i - \langle \hat a\rangle )^2 = 0,  \quad   p_i = \Tr \hat r \hat\pi_i   ,                       \label{eqD.6} 
        \EEQ
which implies, for each $i$, either $p_i = 0$ or $\langle \hat a \rangle=a_i$.  
These equalities are satisfied only if $\langle \hat a \rangle=a_i$ for one specific value of $i$, say $i =1$ 
(so that $\langle \hat a \rangle=a_1$) and if $p_i = 0$ for any $i\neq1$. 
The latter property is equivalent to $p_1= \Tr \hat r \hat \pi_1 =1$ or to $\hat r =\hat \pi_1 \hat r \hat \pi_1$. Hence, starting from the initial state $\hat r$, a measurement of the dispersionless observable $\hat a$ will always produce the outcome $a_1$ and will leave $\hat r$ unchanged.
 
         The identity $\hat r =\hat \pi_1 \hat r \hat \pi_1 $ implies that $\cH_r$ is a subspace of $\cH_1$, so that $1\le  k \le  m_1 \le n$. The considered observable has the form $\hat a = a_1 \hat \pi_1 +\hat b$, where the operator $\hat b$, which belongs to the subspace of $\cH$ complementary to $\cH_1$, is represented by a $(n-m_1) \times (n-m_1)$ matrix. Looking for arbitrary dispersionless observables associated with the state $\hat r$, we take for $m_1$ its smallest possible value $k$. Such observables have therefore the form
\BEQ
\hat a = c \hat \pi_r + \hat b,
\EEQ
where $c$ is an arbitrary real number and $\hat b$ is an arbitrary $(n-k) \times (n-k)$ Hermitean matrix, which depends on $(n-k)^2$ real parameters.

Altogether, whereas the set of all observables depend on $n^2$ parameters, the dispersionless ones depend on $(n-k)^2 +1$ parameters. For mixed states that encompass the whole Hilbert space ($k=n$), e.g., Gibbs equilibrium states, the only dispersionless observables are the trivial ones $c \hat I$. 
In contrast,  pure states $(k=1$) are the least ``noisy'' ones in the sense that their set of dispersionless observables is the largest possible, depending on $n^2-2n +2$ free parameters. For pure states, fluctuations occur for the smallest possible set of observables.

\section*{Acknowledgements}
 We wish to thank Jean-Paul Blaizot, Philippe Grangier, Frank Lalo\"e and anonymous referees for their critical reading and comments. 
 \\
The work of A.E. A. was supported by SCS of Armenia, grants No. 20TTAT-QTa003 and No. 21AG-1C038.


%

\end{document}